\documentclass[stslayout, noinfoline]{imsart}
\usepackage{geometry} 
\usepackage{graphicx}
\usepackage{amssymb, amsmath}
\usepackage{epstopdf}
\usepackage{subfigure}
\usepackage{natbib}
\usepackage[colorlinks,citecolor=blue,urlcolor=blue]{hyperref}
\usepackage[table]{xcolor}
\usepackage{multirow}
\usepackage{tikz}
\usepackage[colorlinks,citecolor=blue,urlcolor=blue]{hyperref}

\usetikzlibrary{intersections, backgrounds, calc, shapes.misc}

\definecolor{light}{RGB}{220, 188, 188}
\definecolor{mid}{RGB}{185, 124, 124}
\definecolor{dark}{RGB}{143, 39, 39}
\definecolor{highlight}{RGB}{0, 255, 0}
\definecolor{gray10}{gray}{0.1}
\definecolor{gray20}{gray}{0.2}
\definecolor{gray30}{gray}{0.3}
\definecolor{gray40}{gray}{0.4}
\definecolor{gray60}{gray}{0.6}
\definecolor{gray70}{gray}{0.7}
\definecolor{gray80}{gray}{0.8}
\definecolor{gray90}{gray}{0.9}
\definecolor{gray95}{gray}{0.95}

\graphicspath{{figures/}}

\begin{document}

\begin{frontmatter}

\title{Calibrating Model-Based Inferences and Decisions}
\runtitle{Model-Based Calibration}

\begin{aug}
  \author{Michael Betancourt%
  \ead[label=e1]{betanalpha@gmail.com}}

  \runauthor{Betancourt}

  \address{Michael Betancourt is a research scientist at 
           Symplectomorphic, LLC. \printead{e1}.}

\end{aug}

\maketitle

\begin{abstract}
As the frontiers of applied statistics progress 
through increasingly complex experiments we must
exploit increasingly sophisticated inferential 
models to analyze the observations we make.  In 
order to avoid misleading or outright erroneous
inferences we then have to be increasingly diligent 
in scrutinizing the consequences of those modeling 
assumptions.  Fortunately model-based methods of 
statistical inference naturally define procedures 
for quantifying the scope of inferential outcomes 
and calibrating corresponding decision making processes.
In this paper I review the construction and 
implementation of the particular procedures that 
arise within frequentist and Bayesian methodologies.
\end{abstract}

\end{frontmatter}

\newpage

\tableofcontents

\newpage

As observations and experiments become more sophisticated,
and we ask correspondingly more detailed questions about
the world around us, we must consider increasingly more
complex inferential models.  The more complex the model,
however, the more subtle the corresponding inferences, 
and the decisions informed by those inferences, will behave.

Consequently understanding how inferences and decisions
vary across the the many possible realizations of a 
measurement becomes a critical aspect in the design and
preliminary evaluation of new observational efforts.  
Such \emph{sensitivity} analyses have a long history in 
the applied sciences but they are often built upon 
heuristics.  Fortunately, formal methods of statistical 
inference naturally admit procedures for understanding
and then calibrating the inferential consequences of 
measurements within the scope of a statistical model.  
The exact mathematical construction of this calibration, 
and the crucial implementation details, depend critically 
on the exact form on inference that we consider.

In this paper I review how inferential outcomes are formally 
calibrated within both the frequentist and Bayesian 
perspectives.  I discuss not only the procedures but also 
the conceptual and practical challenges in implementing these 
procedures in practice, and demonstrate their application 
towards calibrating traditional discovery and limit setting 
results.

\section{Mathematical Preliminaries}

In order to be as explicit as possible when introducing
new functions I will use the conventional mathematical notation.  
A function, $f$, that maps points $x$ in a space $X$ to points 
$y = f(x)$ in a space $Y$ is denoted
\begin{alignat*}{6}
f:\; &X& &\rightarrow& \; &Y&
\\
&x& &\mapsto& &f(x)&.
\end{alignat*}

The real number line will be denoted $\mathbb{R}$ with the 
$N$-dimensional real numbers denoted $\mathbb{R}^{N}$.

Sets of objects are denoted with curly braces, $\{ \ldots \}$,
and a vertical line in between braces denotes a selection
condition which defines a set.  For example,
\begin{equation*}
\{x \in X \mid f(x) = 0 \}
\end{equation*}
defines the subset of points in $X$ that satisfies the
condition $f(x) = 0$.

Finally,
\begin{equation*}
x \sim \pi(x)
\end{equation*}
implies that the space $X \ni x$ is endowed with the probability 
distribution, $\pi$.  If the left-hand side is decorated with a 
tilde, 
\begin{equation*}
\tilde{x} \sim \pi(x) 
\end{equation*}
then this implies that $\tilde{x}$ is a sample from the probability 
distribution $\pi$.

\section{Inference}

Ultimately statistics is a tool to learn about the 
phenomenological behavior of some latent system, for
example the internal structure and dynamics of a subatomic 
particle, the phenotypical encoding of a genome, or 
the response of a population of individuals to a particular 
stimulus.  Although we cannot observe these phenomena 
directly we can probe them through measurements of the 
system, or more precisely measurements of how the system 
interacts with a surrounding environment.  These 
experimental probes can be passive, assembling and 
analyzing data collected for other purposes, or active, 
collecting data from dedicated experiments.

Formally any measurement process defines a 
\emph{measurement space}, $Y$, containing all of the 
possible realizations of a measurement.  These 
realizations, or \emph{observations} are inherently 
stochastic, varying from measurement to measurement.  
If we assume that this variation is sufficiently 
well-behaved, then we can mathematically quantify it
with a \emph{probability distribution} over $Y$.  I 
will refer to any probability distribution over the 
measurement space as a \emph{data generating process}.  
Under this assumption observations of a given system 
are modeled as independent samples from some true data 
generating process, $\pi^{*}$.

\begin{figure*}
\centering
\begin{tikzpicture}[scale=0.35, thick]

  \draw[->, >=stealth, line width=5] 
    (41.5, 5) .. controls (41, 15) and (4.5, 15) .. (4, 7.5);
    
  \draw[->, >=stealth, line width=3, white] 
    (41.5, 5) .. controls (40.9, 15.03) and (4.3, 15.05) .. (4.02, 7.7);
  
  \fill [rounded corners=2pt, color=white] (37.5, 2.5) rectangle (45.5, 5.5);
  \draw [rounded corners=2pt] (37.5, 2.5) rectangle (45.5, 5.5)
  node[midway, align=center] { Inferences and\\Decisions }; 
  
  \draw[->, >=stealth, line width=5] (36, 4) -- (38, 4);
  \draw[->, >=stealth, line width=3, white] (36, 4) -- (37.8, 4);
  
  \draw [rounded corners=2pt] (28, 2.5) rectangle (36, 5.5) 
  node[midway, align=center] { Observation, $\tilde{y}$ }; 
  
  \fill [rounded corners=2pt, fill=dark, text=white] (9.5,0) rectangle (26.5, 10) 
  node [midway, yshift=37, align=center] {Measurement Process};
  
  \draw[->, >=stealth, line width=5] (25.5, 4) -- (28.5, 4);
  \draw[->, >=stealth, line width=3, white] (25.5, 4) -- (28.3, 4);
  
  \fill [rounded corners=2pt, fill=mid, text=white] (10.5, 0.5) rectangle (25.5, 7.5) 
  node[midway, yshift=10, align=center] { Space of observations, $Y$ }
  node[midway, yshift=-10, align=center] { True data generating process, $\pi^{*}$ };

  \draw[->, >=stealth, line width=5] (8, 5) -- (10, 5);
  \draw[->, >=stealth, line width=3, white] (8, 5) -- (9.8, 5);

  \fill [rounded corners=2pt, fill=dark, text=white] (0, 2.5) rectangle (8, 7.5) 
  node [midway, align=center] {Latent System\\Being Studied};
  
\end{tikzpicture}
\caption{Inference is an inductive process that aims to quantify
which phenomenological behaviors of a latent system are consistent 
with measured observations.  Formally we assume that the measurement 
process defines a space of observations, $Y$, and a probability 
distribution quantifying the variation in those observations, 
$\pi^{*}$.  Any realization of the measurement process results in 
an observation, $\tilde{y}$, from which we base our inferences about
the latent system and any resulting decisions about how to interact 
with the latent system.
}
\label{fig:inference_cartoon}
\end{figure*}
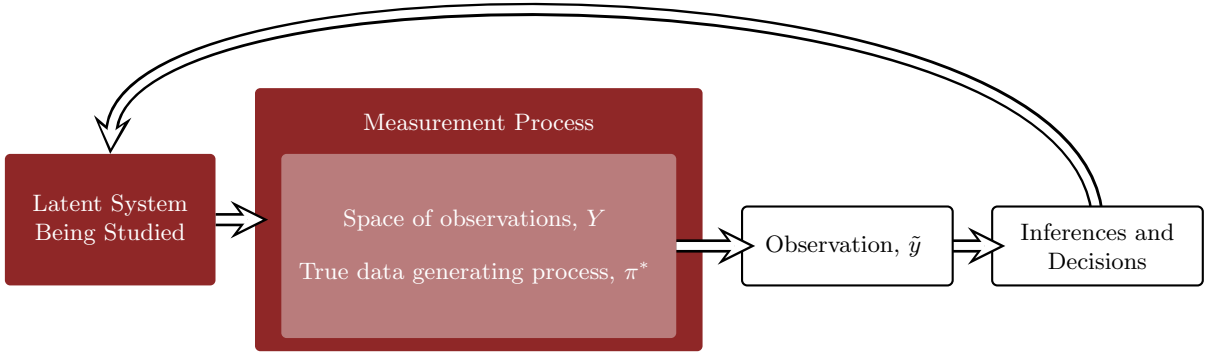

Inference is any procedure that uses observations to 
inform our understanding of the latent system and its 
behaviors (Figure \ref{fig:inference_cartoon}).  
Because of the inherent stochasticity of the measurement 
process, however, any finite observation will convey 
only limited information.  Consequently inference 
fundamentally concerns itself with quantifying the 
\emph{uncertainty} in this understanding.

In particular, there will be many phenomenological 
behaviors consistent with a given observation and, 
in general, those behaviors that appear most consistent 
will vary along with the form of the observation itself.  
Here I will define the \emph{sensitivity} of an experiment 
as the distribution of inferential outcomes induced by 
the variation in the observations.  Ideally our inferences 
would be accurate and capture the true behavior of the 
latent system regardless of the details of particular 
observation, but there are no generic guarantees.  
Consequently in practice we must be careful to study 
these sensitivities with respect to our inferential goals.

Sensitivity analyses become even more important when 
we consider decisions informed by our inferences.
Conventions in many fields focus not on reporting
uncertainties but rather making explicit claims about 
the latent system being studied.  These claims commonly 
take the form of \emph{discovery}, where a particular 
phenomenon is claimed to exist or not exist.   In order 
to limit the possibility that we falsely claim to have 
discovered the presence, or absence, of a phenomenon we 
have to carefully consider the sensitivity of these 
claims.  

Decisions, however, are not always so obvious.  Even
the simple presentation of our inferences requires 
implicit decisions in the form of how we summarize 
and communicate our results.  To ensure that we are
not biasing our audience or ourselves we have to
consider how this presentation would vary with the
underlying observations.

Ultimately, in order to ensure robust analyses we 
have to carefully \textit{calibrate} the consequences
of our inferences.  First consider a set of actions
that we can take, $a \in A$.  Assuming a perfect 
characterization of the latent system of interest we 
could theoretically quantify the relative loss of 
taking a given action with the true \emph{loss function},
\begin{align*}
L^{*} :& \, A \rightarrow \mathbb{R}
\\
& \,\, a \, \mapsto L^{*} (a).
\end{align*}
If we convolving this loss function with an \emph{inferential 
decision-making process} that maps observations to actions,
\begin{align*}
\hat{a} :& \, Y \rightarrow A
\\
& \, y \,\, \mapsto \hat{a}(y),
\end{align*}
we induces a true \emph{inferential loss function},
\begin{align*}
L^{*}_{\hat{a}} :& \, Y \rightarrow \mathbb{R}
\\
& \, y \,\, \mapsto L^{*} (\hat{a}(y)).
\end{align*}
More optimistic readers might also consider the
equivalent utility function, $U^{*}(a) \equiv - L^{*}(a)$.

A sensitivity analysis considers the distribution of 
$L^{*}_{\hat{a}} (y)$ with varying observations, $y$, 
while a calibration considers expected values of the 
loss function over the possible observations.  For 
example, we might aim to calibrate an experiment to 
ensure that the average loss is below a certain value 
or that a particular quantile of the loss distribution 
is below a certain value.

These analyses, however, require that we know both
the true behavior of the latent system, so that we 
can quantify the relative loss of each action, and the 
true data generating process, so that we can quantify 
the variation of possible observations.  In practice 
we don't know the true nature of the latent system or 
the true data generating process of the resulting 
measurements, but we can \emph{model} them.  
\emph{Statistical models} quantify the scope of possible 
data generating processes, allowing us to quantify the 
sensitivity of inferences and then construct formal 
calibrations within that scope.

\section{Model-Based Sensitivities and Calibration}

Inferences that explicitly model the measurement process, 
or \emph{model-based inferences}, naturally define the 
scope of the possible observations, leaving practitioners 
to employ their domain expertise to construct relevant 
loss functions.  

In this section I review the model configuration space
that underlies model-based inference and how the 
frequentist and Bayesian paradigms utilize this space 
to define inference and calibrate inferential outcomes.

\subsection{The Model Configuration Space}

\begin{figure*}
\centering
\subfigure[]{
\begin{tikzpicture}[scale=0.225, thick]

  \draw[color=white] (-15, 0) -- (15, 0);

  \fill[mid] (0, 0) ellipse (13 and 7);
  \node at (12, -6) {$\mathcal{P}_{Y}$};
  
  \fill[color=white] (-7, 3) circle (6pt)
  node[right, color=white] {$\pi^{*}$};
  
\end{tikzpicture}
}
\subfigure[]{
\begin{tikzpicture}[scale=0.225, thick]
  
  \draw[color=white] (-15, 0) -- (15, 0);
  
  \fill[mid] (0, 0) ellipse (13 and 7);
  \node at (12, -6) {$\mathcal{P}_{Y}$};
  
  \fill[dark] (-4, 2) ellipse (6 and 3);
  \node[color=white] at (1.5, -0.5) {$\mathcal{M}$};
  
  \fill[color=white] (-7, 3) circle (6pt)
  node[right, color=white] {$\pi^{*}$};
  
\end{tikzpicture}
}
\caption{(a) The true data generating process, $\pi^{*}$,
belongs in the space of all possible data generating
processes over the measurement space, $\mathcal{P}_{Y}$.
(b) Model-based inferences consider a model configuration
space, $\mathcal{M}$, which contains only a limited subset
of all of those possible data generating processes. 
}
\label{fig:big_and_small_worlds}
\end{figure*}
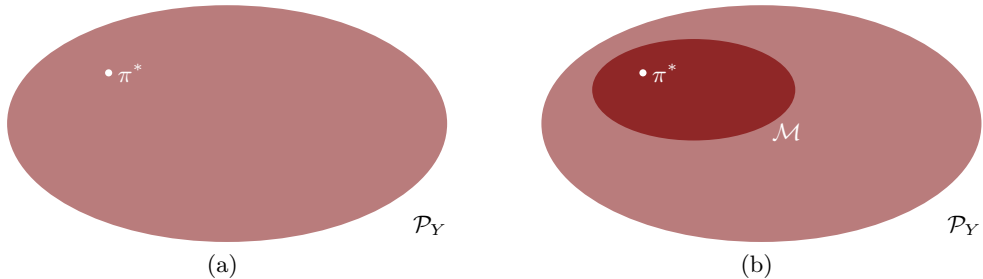

A statistical model establishes a \emph{model 
configuration space}, or a collection of data generating 
processes.  Ideally the model configuration space is 
designed to be sufficiently rich to either contain the 
true data generating process or, more realistically, 
contain data generating processes sufficiently similar 
to the true data generating process within the resolution
of our experimental probe (Figure \ref{fig:big_and_small_worlds}).  
Each individual data generating process in the model 
configuration space will be denoted a \emph{model configuration}.

Typically the model configuration space admits a 
parameterization in which each model configuration can be 
identified with a parameter value, $\theta \in \Theta$. 
Often these parameters decompose into components,
\begin{equation*}
\theta = \left\{\theta_{1}, \ldots, \theta_{N}\right\} 
\end{equation*}
with each component responsible for quantifying only one 
aspect of the phenomenological behavior of the latent 
system, the environment containing the latent system, or 
the measurement process itself.  Such parameter 
decompositions make the model configuration space 
particularly interpretable.

Moreover, if each model configuration admits a density 
with respect to a common reference measure over $Y$ 
then we can fully specify the model configuration space 
with the family of densities $\pi (y \mid \theta)$ for 
$y \in Y$ and $\theta \in \Theta$.  In many applied 
fields this family of densities, or even the model 
configuration space itself, is introduced as the 
\emph{likelihood}.  Unfortunately, that term has a 
more precise definition in the statistics literature:
the phrase likelihood is used exclusively to denote the 
function over the parameter space given by evaluating 
each density at a particular measurement $\tilde{y} \in Y$,
\begin{align*}
\mathcal{L}_{\tilde{y}} &: \Theta \rightarrow \mathbb{R} 
\\
& \quad \theta \, \mapsto \pi(\tilde{y} \mid \theta).
\end{align*}
Consequently I will avoid the use of likelihood unless 
the model configuration densities are explicitly being 
evaluated at a given observation.

\begin{figure*}
\centering
\begin{tikzpicture}[scale=0.35, thick]

  \draw[->, >=stealth, line width=5] 
    (41.5, 9) .. controls (41, 19) and (22, 19) .. (20, 16);
    
  \draw[->, >=stealth, line width=3, white] 
    (41.5, 9) .. controls (40.9, 19.1) and (22, 19.05) .. (20.1, 16.15);
  
  \draw[->, >=stealth, line width=5] 
    (15, 16) .. controls (15, 19) and (4, 19) .. (4, 7.5);
    
  \draw[->, >=stealth, line width=3, white] 
    (15, 16) .. controls (15, 19) and (4, 19.1) .. (4, 7.7);

  \fill [rounded corners=2pt, color=white] (37.5, 6.5) rectangle (45.5, 9.5);
  \draw [rounded corners=2pt] (37.5, 6.5) rectangle (45.5, 9.5) 
  node[midway, align=center] { Inferences and\\Decisions }; 
  
  \draw[->, >=stealth, line width=5] (35, 4) -- (38, 7);
  \draw[->, >=stealth, line width=3, white] (35, 4) -- (37.85, 6.85);
  
  \draw [rounded corners=2pt, fill=white] (28, 2.5) rectangle (36, 5.5) 
  node[midway, align=center] { Observation, $\tilde{y}$ }; 
  
  \draw[->, >=stealth, line width=5] (25, 13.5) -- (38, 9);
  \draw[->, >=stealth, line width=3, white] (25, 13.5) -- (37.8, 9.08);
  
  \fill [rounded corners=2pt, fill=mid, text=white] (10.5, 11) rectangle (25.5, 16) 
  node[midway, align=center] 
  { Model Configuration Space,\\$\pi(y \mid \theta), \,\theta \in \Theta$ };
  
  \fill [rounded corners=2pt, fill=dark, text=white] (9.5,0) rectangle (26.5, 10) 
  node [midway, yshift=37, align=center] {Measurement Process};
  
  \draw[->, >=stealth, line width=5] (25.5, 4) -- (28.5, 4);
  \draw[->, >=stealth, line width=3, white] (25.5, 4) -- (28.3, 4);
  
  \fill [rounded corners=2pt, fill=mid, text=white] (10.5, 0.5) rectangle (25.5, 7.5) 
  node[midway, yshift=10, align=center] { Space of observations, $Y$ }
  node[midway, yshift=-10, align=center] { True data generating process, $\pi^{*}$ };

  \draw[->, >=stealth, line width=5] (8, 5) -- (10, 5);
  \draw[->, >=stealth, line width=3, white] (8, 5) -- (9.8, 5);

  \fill [rounded corners=2pt, fill=dark, text=white] (0,2.5) rectangle (8, 7.5) 
  node [midway, align=center] {Latent System\\Being Studied};
  
\end{tikzpicture}
\caption{In model based inference the measurement process is
modeled with a model configuration space, $\Theta$.  Inferences
identify which model configurations, $\theta \in \Theta$, are
consistent with a given observation, which then informs the
properties of the true data generating process and the latent
system under investigation.
}
\label{fig:inference}
\end{figure*}
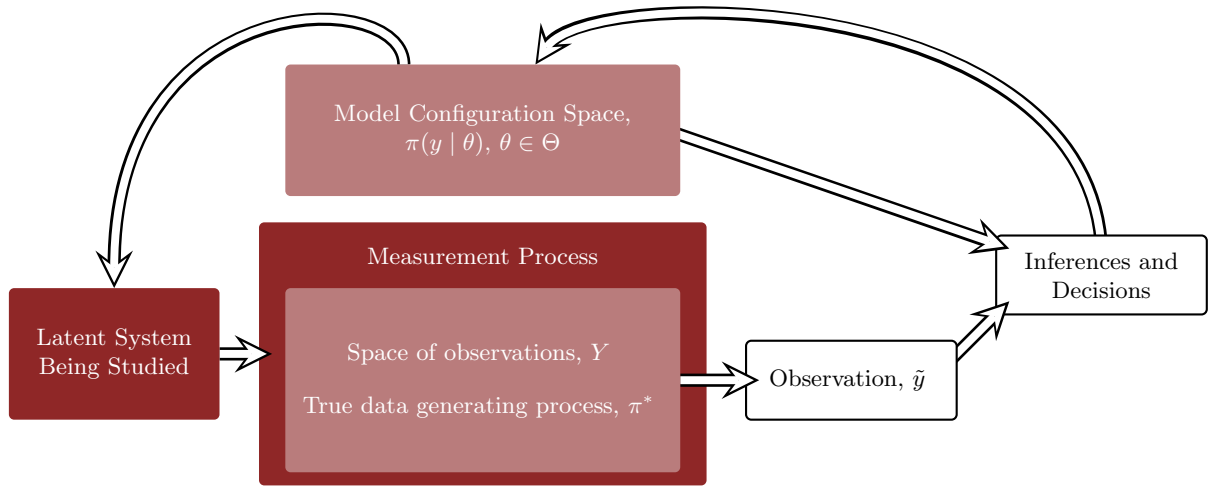

Once we have constructed a model configuration space, 
inference becomes a means of identifying those model 
configurations that are consistent with an observation
(Figure \ref{fig:inference}).  Because there is no 
unique definition of this sort of consistency, there 
are multiple approaches to inference.  The two employed 
most in practice are \emph{frequentist} and \emph{Bayesian} 
inference.

\subsection{Model-Based Loss Functions}

Given a means of quantifying those model configurations
consistent with an observation, we can use that 
quantification to motivate fruitful decisions about how to 
best interact with the latent system. For example we may 
want to intervene with the system or systems like it, 
introducing a treatment or altering the surrounding 
environment.  Alternatively we may want to decide on 
whether or not to claim the discovery of the absence or 
presence of a new phenomenon, or whether or not to follow 
up with another experiment.

For model-based inference we can quantify the utility
of a set of actions by defining a loss function for
each possible data generating process in our model,
\begin{align*}
L :& \, A \times \Theta \rightarrow \mathbb{R}
\\
& \,\,\, (a, \theta) \,\, \mapsto L(a, \theta).
\end{align*}
Given an inferential decision making process
\begin{align*}
\hat{a} :& \, Y \rightarrow A
\\
& \, y \,\, \mapsto \hat{a}(y),
\end{align*}
we can then define a model-based inferential loss function,
\begin{align*}
L_{\hat{a}} :& \, Y \times \Theta \rightarrow \mathbb{R}
\\
& \,\,\, (y, \theta) \,\, \mapsto L(\hat{a}(y), \theta).
\end{align*}
Presuming that the true data generating process is 
contained within the model configuration space, 
sensitivity analyses and calibration within the 
scope of our model quantifies the actual performance 
of our decisions.  When the true data generating
process is close to the model configuration space
then this process approximates the actual performance.

That said, constructing a loss function alone is 
insufficient to admit sensitivity analyses and
calibrations.  Before the measurement we are 
ignorant of not only what the observation will be, 
and hence what action we will take, but also which 
model configuration gives the true data generating 
process.  In order to quantify the performance of 
our decisions we have to define the scope of possible 
observations and possible data generating processes.  
Exactly how we do that depends intimately on the 
formal details of the inferences that we make.

\subsection{Sensitivity and Calibration of Frequentist Inference}

Frequentist inference \citep{CasellaEtAl:2002, 
Lehmann:2006, Keener:2011} derives from the interpretation
that probability theory can model only the frequencies of
repeatable processes.  This definition is consistent with 
the use of probabilities to model the inherent variation 
of observations, but it does not allow us to define 
probabilities over the model configuration space itself, 
as those probabilities would not correspond to the 
hypothetical frequencies of any repeatable process.  
Ultimately this strong philosophical assumption implies 
that we cannot use any form of weighting to quantify 
consistency in the model configuration space because any 
self-consistent weighting is equivalent to the assignment 
of probabilities!

Consequently frequentist inference must take the form of 
definite decisions about which parts of the model 
configuration space are consistent with an observation and 
which are not.  From a frequentist perspective inference 
and decisions are one in the same!  Because such definite 
decisions can readily exclude the true data generating 
process, or model configurations close to the true data 
generating process, from consideration we have to carefully 
calibrate these decisions so that such exclusions are 
sufficiently rare.  

Ultimately frequentist inference does not define exactly
how an observation informs which parts of the model 
configuration space to keep and which to discard.  Rather
frequentist inference establishes a means of calibrating 
any such procedure that might be considered.

\subsubsection{Frequentist Inference}

Any procedure that consumes an observation to produce
a definite decision about which parts of the model 
configuration space are considered consistent takes
the mathematical form of an \emph{estimator}.  Estimators 
are functions from the measurement space to subsets of 
the model configuration space, mapping observations to
subsets of model configurations,
\begin{equation*}
\hat{\theta} : Y \rightarrow \mathcal{T},
\end{equation*}
where $\mathcal{T}$ is the space of well-defined subsets of 
the model configuration space, $\Theta$.

\begin{figure*}
\centering
\subfigure[]{
\begin{tikzpicture}[scale=0.225, thick]

  \draw[color=white] (-15, 0) -- (15, 0);

  \fill[mid] (0, 0) ellipse (13 and 7);
  \node at (12, -6) {$\mathcal{P}_{Y}$};
  
  \draw[color=dark, dashed] (-4, 2) ellipse (6 and 3);
  \node[color=white] at (1.5, -0.5) {$\mathcal{M}$};
  
  \fill[color=dark] (-3, 1.75) circle (6pt)
  node[right, color=white] {$\hat{\theta}(\tilde{y})$};
  
  \fill[color=white] (-7, 3) circle (6pt)
  node[right, color=white] {$\pi^{*}$};
  
\end{tikzpicture}
}
\subfigure[]{
\begin{tikzpicture}[scale=0.225, thick]

  \draw[color=white] (-15, 0) -- (15, 0);

  \fill[mid] (0, 0) ellipse (13 and 7);
  \node at (12, -6) {$\mathcal{P}_{Y}$};
  
  \draw[color=dark, dashed] (-4, 2) ellipse (6 and 3);
  \node[color=white] at (1.5, -0.5) {$\mathcal{M}$};
  
  \fill[color=dark] (-2.75, 2) ellipse ({4} and {2.5})
  node[color=white] {$\hat{\theta}(\tilde{y})$};
  
  \fill[color=white] (-7, 3) circle (8pt)
  node[right, color=white] {$\pi^{*}$};
  
\end{tikzpicture}
}
\caption{(a) Point estimators identify a single model 
configuration, $\hat{\theta}(\tilde{y})$, that is 
ideally close to the true data generating process, 
$\theta^{*}$, for any given observation $\tilde{y}$.  
(b) Confidence sets identify entire subsets of the 
model configuration space that ideally contain the
true data generating process for any given observation.
}
\label{fig:frequentist_inference}
\end{figure*}
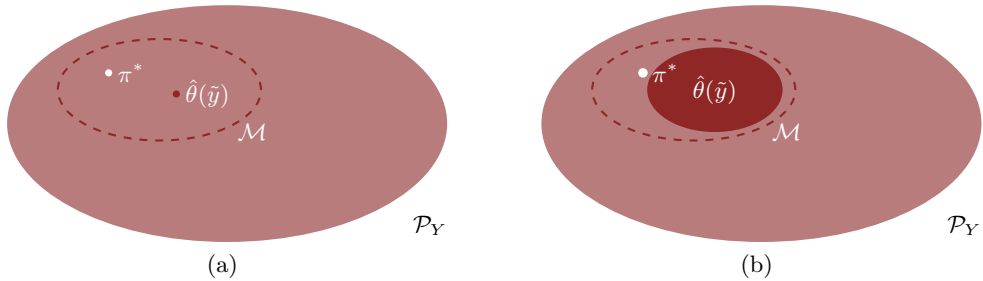

A common class of estimators are \emph{point estimators} 
that identify a single point in the model configuration space
(Figure \ref{fig:frequentist_inference}a),
\begin{equation*}
\hat{\theta} : Y \rightarrow \Theta,
\end{equation*}
Point estimators formalize the intuition of a ``best fit'',
where inferences are summarized with a single point at the
expense of ignoring the uncertainty inherent in learning 
from finite observations.

The more general class of estimators that identify entire 
subsets of the model configuration space are known as
\emph{confidence sets} (Figure 
\ref{fig:frequentist_inference}b), or \emph{confidence 
intervals} if the model configuration space is 
one-dimensional.  The nomenclature is meant to suggest 
that if a confidence set has been properly constructed 
then we can be confident that these sets will contain 
the true data generating process for sufficiently many 
observations.

\subsubsection{The Frequentist Calibration Criterion}

The actual choice of which estimator to employ in a given
analysis is left to the practitioner.  Constraints that 
enforce desired properties can be imposed to restrict the 
space of potential estimators, but the choice of these
desired properties remains the responsibility of the
practitioner.  Regardless of how we ultimately select an 
estimator, however, we can use the model configuration 
space to calibrate the estimator and determine it's 
practical utility.

In frequentist inference our actions are definite 
quantifications of the model configuration space and,
by construction, estimators are inferential decision
making processes.  In order to define a calibration
criterion we we must first construct a model-based 
loss function,
\begin{equation*}
L (\hat{\theta}, \theta),
\end{equation*}
that quantifies how well $\hat{\theta}$ identifies 
the true data generating process, $\theta$. Substituting
an estimator yields the model-based inferential loss 
function, 
\begin{equation*}
L_{\hat{\theta}} (y, \theta) 
\equiv 
L(\hat{\theta}(y), \theta). 
\end{equation*}
As with estimators there is no canonical loss function 
in frequentist inference; instead one must be chosen 
using whatever domain expertise is available within 
the context of a particular analysis.  In practice this 
choice often considers the structure of the estimator 
itself.

If we knew that a given $\theta$ identified the true 
data generating process then the sensitivity of the loss 
of the estimator over the possible observations could be 
summarized with an expectation over that configuration. 
This expectation yields an \emph{expected loss} for each 
model configuration,
\begin{equation*}
L_{\hat{\theta}} (\theta) 
= 
\int_{Y} \mathrm{d} y \, \pi(y \mid \theta) \, 
L_{\hat{\theta}} (y, \theta).
\end{equation*}

Because we don't have any information about which model
configuration identifies the true data generating process
before a measurement is made, the frequentist calibration 
criterion is defined as the maximum expected loss over all 
possible model configurations,
\begin{equation*}
\bar{L}_{\hat{\theta}}
= 
\max_{\theta} L_{\hat{\theta}} (\theta).
\end{equation*}
If the model configuration space is sufficiently rich 
that it contains the true data generating process, then 
this calibration criterion defines the worst case loss 
of our given estimator.  Bounding the worst case loss 
of an estimator is an extremely powerful guarantee on 
its practical performance, but also a very conservative 
one as bounds can be dominated by unrealistic but not
impossible data generating processes towards the 
boundaries of the model configuration space.

A natural loss function for point estimators is the 
$L^{p}$ distance between the estimated model 
configuration and the presumed true data generating 
process,
\begin{equation*}
L^{p}_{\hat{\theta}} (y, \theta) 
= 
( ( \theta - \hat{\theta} (y) )^{2} )^{p/2}.
\end{equation*}
The expected $L^{p = 2}$ loss is known as the  
\emph{variance} of an estimator.  

Similarly, a natural natural loss function for confidence
sets is inclusion of the presumed true data generating 
process,
\begin{equation*}
L_{\hat{\theta}} (y, \theta) 
= 
\mathbb{I}_{\theta} [ \hat{\theta}(y) ],
\end{equation*}
where the indicator function, $\mathbb{I}$, is defined as
\begin{equation*}
\mathbb{I}_{\theta} [ T ] =
\left\{
\begin{array}{rc}
1, & \theta \in T \\
0, & \mathrm{else}
\end{array} 
\right. .
\end{equation*}
The expected inclusion loss, or \emph{coverage} is simply 
how often the confidence set contains the presumed true 
data generating process. 

While this calibration procedure can used to analyze the
frequentist properties of a given estimator, they can also 
be used to optimize the choice of estimator.  Given a 
family of estimators, $\{ \hat{\theta}_{x} \}$, the optimal 
estimator will satisfy the \emph{minimax} criterion,
\begin{align*}
\hat{\theta}^{*} 
&= \underset{x}{\mathrm{argmin}} \, L_{\hat{\theta}_{x}}^{*}
\\
&= \underset{x}{\mathrm{argmin}} \max_{\theta} \, 
L_{\hat{\theta}_{x}} (\theta)
\\
&= \underset{x}{\mathrm{argmin}} \max_{\theta} \,
\int_{Y} \mathrm{d} y \, \pi(y \mid \theta) \, 
L_{\hat{\theta}_{x}} (y, \theta).
\end{align*}
For example, a desired coverage might be established 
initially and then a confidence set engineered to 
ensure that the coverage is met, or exceeded, for all 
of the data generating processes in the model 
configuration space.

\subsubsection{Frequentist Methods in Practice}

Aside from the conceptual challenge of choosing a
loss function that enforces the needs of a given 
analysis, the computational burden of frequentist
calibration is a significant impediment to its 
application.  In particular, even approximately 
scanning through the model configuration space to 
identify the maximal expected loss often requires
more computational resources than realistically
available to a practitioner.

Many frequentist analyses assume sufficiently simple 
model configuration spaces, estimators, and loss 
functions such that the maximum expected loss can be 
computed analytically.  The analytic results allow, for 
example, optimal estimators to be chosen from families 
of candidate estimators with strong guarantees on the 
performance of the best choice.  The practical validity 
of these guarantees, however, requires that the true 
data generating process be simple enough that it can be 
contained within the relatively crude model configuration 
space.  For the complex experiments of applied interest 
this can be a dangerous assumption.

Without analytic results one might consider interpolative 
methods that bound the variation in the expected loss 
between a grid of points distributed across the model 
configuration space.  At each of these points Monte Carlo 
methods can be used to simulate observations and approximate
the expected loss, and then the properties of the loss
function itself can be used to interpolate the expected
loss amidst the grid points.  These methods can yield 
reasonable results for low-dimensional model configuration
spaces, but as the dimensionality of the model increases 
even strong smoothness assumptions can become insufficient 
to inform how to interpolate between the grid points.

In order to avoid this curse of dimensionality frequentist 
analyses unaccommodating to analytic results often resort 
to \emph{asymptotics}. Asymptotic analyses assume that 
the model configuration space is sufficiently regular that 
as we consider more observations at once the behavior of 
the model configuration space follows a \emph{central limit 
theorem}. Under these conditions the likelihood for any 
observation concentrates in an increasingly small 
neighborhood around the \emph{maximum likelihood estimator},
\begin{equation*}
\theta_{\mathrm{ML}} (y) =
\underset{\theta \in \Theta}{\mathrm{argmax}} \,
\pi ( \tilde{y} \mid \theta).
\end{equation*}
Moreover, in this limit the breadth of that neighborhood 
is given by the inverse of the Fisher information matrix,
\begin{equation*}
I(\tilde{y}) = 
\left. \nabla^{2} \pi ( \tilde{y} \mid \theta) 
\right|_{\theta = \theta_{\mathrm{ML}} (y) }.
\end{equation*}
The concentration in the model configuration space 
in this asymptotic limit admits convenient analytic 
approximations to the frequentist calibration procedure. 

Asymptotic behavior also motivates the concept of 
\emph{profiling}, which is of use when the parameter
space separates into phenomenological parameters
related to the underlying system of interest and 
\emph{nuisance} or \emph{systematic} parameters 
that are unrelated to that system but still effect 
the data generating process.  Under certain conditions 
the observations inform the nuisance parameters
faster than the the phenomenological parameters;
in the asymptotic limit the uncertainty in these 
parameters becomes negligible and they can be 
replaced with conditional maximum likelihood estimates.  

More formally, if the parameterization of the model
configuration space decomposes into phenomenological 
parameters, $\vartheta$, and nuisance parameters,
$\sigma$, then we define the \emph{conditional maximum
likelihood estimator} as
\begin{equation*}
\hat{\sigma}(\vartheta, \tilde{y})
=
\underset{\sigma}{\mathrm{argmax}} \,
\pi(\tilde{y} \mid \vartheta, \sigma)
\end{equation*}
and the corresponding \emph{profile likelihood} as
\begin{equation*}
\hat{\pi} ( \tilde{y} \mid \vartheta )
=
\pi ( \tilde{y} \mid \vartheta, 
\hat{\sigma}(\vartheta, \tilde{y}) ).
\end{equation*}
The profile likelihood can then be used to calibrate
estimators of the phenomenological parameters, at
least in this limit.

The utility of these asymptotic methods depends 
critically on the structure of the model 
configuration space and its behavior as we consider 
more observations.  Simpler models typically converge 
to the asymptotic limit faster and hence require fewer 
data for asymptotic calibrations to be reasonably 
accurate.  More complex models, however, converge more 
slowly and may require more data than is practical, or 
they may not satisfy the necessary conditions to 
converge at all.  

Consequently it is crucial to explicitly verify that 
the asymptotic regime has been reached in a given 
analysis.  As with analytic methods, one has to be 
especially careful to not employ an over-simplistic 
model to facilitate the applicability of the asymptotic 
results while compromising the practical validity of 
the resulting calibration.

\subsection{Sensitivity and Calibration of Bayesian Inference}

Bayesian inference \citep{BernardoEtAl:2009, GelmanEtAl:2014a} 
broadens the interpretation of probability theory, allowing 
it to be used to not only model inherent variation in 
observations but also provide a probabilistic quantification 
of consistency between the data generating processes in the 
model configuration space and observations.  

This generalization manifests in a unique procedure for 
constructing  inferences which can then be used to inform 
decisions.  Ultimately Bayesian inference decouples inference 
from decision making, making the assumptions underlying both 
more explicit and often easier to communicate.  Moreover, 
the fully probabilistic treatment of the Bayesian perspective 
immediate defines a procedure for constructing sensitivities 
and calibrations.

\subsubsection{Bayesian Inference}

Bayesian inference compliments the data generating 
processes in the model configuration  space with a 
\emph{prior distribution} over the model configuration 
space itself.  The prior distribution quantifies any 
information on which model configurations are closer 
to the true data generating process than others that
is available before a measurement is made.  This 
information can come from, for example, physical 
considerations, previous experiments, or even expert 
elicitation.  Careful choices of the prior distribution 
can go a long way towards regularizing unwelcome 
behavior of the model configuration space.

\begin{figure*}
\centering
\subfigure[]{
\begin{tikzpicture}[scale=0.225, thick]

  \draw[color=white] (-15, 0) -- (15, 0);

  \fill[mid] (0, 0) ellipse (13 and 7);
  \node at (12, -6) {$\mathcal{P}_{Y}$};
  
  \draw[color=dark, dashed] (-4, 2) ellipse (6 and 3);
  \node[color=white] at (1.5, -0.5) {$\mathcal{M}$};
  
  \begin{scope}
    \clip (-4, 2) ellipse (6 and 3);
    \foreach \i in {0, 0.025,..., 1} {
      \fill[opacity={exp(-5 * \i*\i)}, dark] 
      (-4, 2) ellipse ({7 * \i} and {3 * \i});      
    }
  \end{scope}
  
  \fill[color=white] (-7, 3) circle (8pt)
  node[right, color=white] {$\pi^{*}$};
  
\end{tikzpicture}
}
\subfigure[]{
\begin{tikzpicture}[scale=0.225, thick]

  \draw[color=white] (-15, 0) -- (15, 0);

  \fill[mid] (0, 0) ellipse (13 and 7);
  \node at (12, -6) {$\mathcal{P}_{Y}$};
  
  \draw[color=dark, dashed] (-4, 2) ellipse (6 and 3);
  \node[color=white] at (1.5, -0.5) {$\mathcal{M}$};
  
  \begin{scope}
    \clip (-4, 2) ellipse (6 and 3);
    \foreach \i in {0, 0.05,..., 1} {
      \fill[opacity={exp(-5 * \i*\i)}, dark] 
      (-5, 2.5) ellipse ({4 * \i} and {2 * \i});      
    }
  \end{scope}
  
  \fill[color=white] (-7, 3) circle (8pt)
  node[right, color=white] {$\pi^{*}$};
  
\end{tikzpicture}
}
\caption{(a) Bayesian inference begins with a prior 
distribution, shown here in dark red, over the model 
configuration space, $\mathcal{M}$, that quantifies
information available before a measurement.  (b) 
Information encoded in an observation updates the
prior distribution into a posterior distribution
that ideally concentrates around the true data
generating process, $\pi^{*}$.
}
\label{fig:bayesian_inference}
\end{figure*}
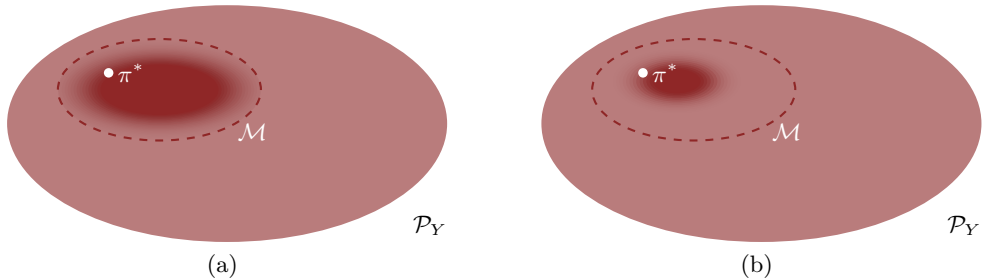

Together the model configuration space and the prior
distribution define the \emph{Bayesian joint distribution} 
over the measurement space and the parameter space,
\begin{equation*}
\pi (y, \theta) = \pi(y \mid \theta) \pi(\theta).
\end{equation*}
The titular Bayes' Theorem conditions this joint
distribution on an observation, $\tilde{y}$, to give 
the \emph{posterior distribution},
\begin{equation*}
\pi( \theta \mid \tilde{y} ) 
=
\frac{\pi( \tilde{y}, \theta) }{ \pi(\tilde{y}) }
\propto
\pi( \tilde{y} \mid \theta ) \, \pi( \theta).
\end{equation*}
In words, the prior distribution quantifies information
available before the measurement, the model configuration 
space decodes the information within an observations, 
and the posterior distribution combines both sources of 
information to quantify the information about the latent
system being studies after a measurement 
(Figure \ref{fig:bayesian_inference}).

Any well-posed statistical query we might make of our 
system reduces to interrogations of the posterior 
distribution.  Mathematically this must take the form 
of a posterior expectation for some function, $f$,
\begin{equation*}
\mathbb{E} [f] = \int_{\Theta} \mathrm{d} \theta \,
\pi( \theta \mid \tilde{y} ) \, f (\theta).
\end{equation*}
For example, we might consider the posterior mean or 
median to identify where the posterior is concentrating 
or the posterior standard deviation or the posterior
quartiles to quantify the breadth of the distribution.

Posterior expectations also define a unique decision 
making process in Bayesian inference.  First we define 
the expected loss for a given action by averaging a 
model-based loss function, $L(a, \theta)$, over the 
posterior distribution,
\begin{equation*}
L(a, \tilde{y}) = \int_{\Theta} \mathrm{d} \theta \,
\pi( \theta \mid \tilde{y} ) L(a, \theta).
\end{equation*}
We can then define a decision making process by taking 
the action with the smallest expected loss, 
\begin{equation*}
a^{*} (\tilde{y}) = \min_{a \in A} L(a, \tilde{y}).
\end{equation*}

For example, our decision might be to summarize the
posterior with a single ``best fit'' model configuration, 
$\hat{\theta}$.  Given the loss function 
\begin{equation*}
L(\hat{\theta}, \theta) 
= 
(\hat{\theta} - \theta )^{2}
\end{equation*}
the expected losses for each possible summary becomes
\begin{equation*}
L(\hat{\theta}, \tilde{y}) = ( \hat{\theta} - \mu( \tilde{y} ) )
+ \sigma^{2} (\tilde{y}),
\end{equation*}
where $\mu( \tilde{y} )$ is the posterior mean and
$\sigma( \tilde{y} )$ is the posterior standard deviation.
Following the Bayesian decision making process, our optimal
decision is to summarize our posterior by reporting the 
posterior mean, 
$\hat{\theta}^{*} (\tilde{y}) = \mu(\tilde{y})$.

\subsubsection{The Bayesian Calibration Criterion}

Bayes' Theorem provides a unique procedure for constructing
inferences and making subsequent decisions given an observation, 
but there are no guarantees that these decisions will achieve 
any desired performance for any possible observation.  
Consequently sensitivity analysis and calibration of this 
decision making process across possible is still important 
in Bayesian inference.

Instead of having to consider each model configuration
equally, however, the prior distribution allows us to diminish 
the effect of unrealistic but not impossible model configurations. 
In particular sampling from the joint distribution generates
an ensemble of reasonable data generating process and 
corresponding observations which we can use to quantify the
performance of our decisions.

For example, we can quantify the sensitivity of any inferential 
outcome by integrating the model configurations out of the 
Bayesian joint to give the \emph{prior data generating process},
\begin{equation*}
\pi ( y ) 
= 
\int_{\Theta} \mathrm{d} \theta \,
\pi( y, \theta)
= 
\int_{\Theta} \mathrm{d} \theta \,
\pi( y \mid \theta ) \, \pi (\theta).
\end{equation*}
The prior data generating process probabilistically 
aggregates the behaviors of all of the possible data 
generating process in the model configuration space 
into a single probability distribution over the 
measurement space.  We can then analyze the sensitivity 
of any inferential outcome by running our analysis over 
an ensemble of observations sampled from this distribution.

Moreover, we can calibrate a decision making process
by integrating a model-based loss function against the 
full Bayesian joint distribution,
\begin{equation*}
\bar{L}_{A} 
= 
\int_{Y, \Theta} \mathrm{d} y \, \mathrm{d} \theta \,
L_{\hat{a}} (y, \theta) \pi(\theta, y).
\end{equation*}
This calibration immediately quantifies the expected 
loss as both the observations and data generating 
processes vary within the scope of our model.

\subsubsection{Bayesian Methods in Practice}

The unified probabilistic treatment of Bayesian inference
ensures that all calculations take the form of expectation
values with respect the Bayesian joint distribution, its
marginals, such as the prior distribution and the prior 
data generating process, or its conditionals, such as the 
posterior distribution.  Consequently calculating expectation 
values, or more realistically accurately estimating them, 
is the sole computational burden of Bayesian inference. 

Posterior expectations are challenging to compute, and
indeed much of the effort on the frontiers of statistical 
research concerns the development and understanding of 
approximation methods.  One of the most powerful and
well-understood of these is Markov chain Monte Carlo
\citep{RobertEtAl:1999, BrooksEtAl:2011} and its state of 
the art implementations like Hamiltonian Monte Carlo
\citep{Betancourt:2017a}.

On the other hand, expectations with respect to the Bayesian 
joint distribution are often amenable to much simpler Monte 
Carlo methods.  In particular, if we can draw exact samples 
from the prior distribution and each of the data generating 
processes in the model configuration space then we can generate 
joint samples with the sequential sampling scheme
\begin{align*}
\tilde{\theta} &\sim \pi(\theta)
\\
\tilde{y} &\sim \pi(y \mid \tilde{\theta}).
\end{align*}
For each simulated observation, $\tilde{y}$, we can
construct a subsequent posterior distribution, make 
posterior-informed decisions, and then compare those
decisions to the simulated truth, $\tilde{\theta}$.  
As we generate a larger sample from the Bayesian joint 
distribution we can more accurately quantify our 
sensitivities and calibrations.  

We can also quantify how sensitivity a calibration 
is to a particular component of the parameter space,
$\vartheta$, by sampling the complementary parameters,
$\sigma$, from the corresponding conditional prior 
distribution,
\begin{align*}
\tilde{\sigma} &\sim \pi(\sigma \mid \vartheta)
\\
\tilde{y} &\sim \pi(y \mid (\vartheta, \tilde{\sigma})).
\end{align*}
This allows us, for example, to see how our decision
making process behaves as for various phenomenlogical
behaviors.

Interestingly, the application of Monte Carlo to the 
Bayesian joint distribution is not at all dissimilar 
to many of the heuristic schemes common in the sciences.  
Sampling $\tilde{y} \sim \pi(y \mid \tilde{\theta})$ 
just simulates the experiment conditioned on the model 
configuration, $\tilde{\theta}$.  The addition step 
$\theta \sim \pi(\theta)$ simply simulates model 
configurations consistent with the given prior information 
instead of selecting a few model configuration by hand.

One inferential outcome immediately amenable to calibration 
is the approximation of posterior expectations themselves. 
\cite{CookEtAl:2006}, for example, introduce a natural
way to calibrate the estimation of any posterior quantiles.
This then immediately provides a procedure for quantifying 
the accuracy of any algorithm that yields deterministic 
approximations to posterior quantiles, for example as
demonstrated in \cite{YaoEtAl:2018}. 
%Similarly, 
%\textbf{Talts et al.} demonstrate how the Bayesian joint
%distribution defines a natural calibration of sampling
%based algorithms, such as Monte Carlo and Markov chain 
%Monte Carlo.

Bayesian sensitivity analysis is particularly useful
for identifying known pathologies in Bayesian inference
by carefully examining the simulated analyses. Consider, 
for example, the \emph{posterior $z$-score} for the 
parameter component, $\tilde{\theta}_{n}$,
\begin{equation*}
z_{n}
= 
\left| \frac{\mu_{n} (\tilde{y})
- \tilde{\theta}_{n}}{ \sigma_{n} (\tilde{y})} \right|,
\end{equation*}
where $\mu_{n} (\tilde{y})$ denotes the posterior mean 
of $\tilde{\theta}_{n}$ and $\sigma_{n} (\tilde{y})$ the
corresponding posterior standard deviation.  The posterior
$z$-score quantifies how much the posterior distribution
envelops the presumed true data generating process along
this direction in parameter space.  At the same time
consider the \emph{posterior shrinkage} of that parameter 
component,
\begin{equation*}
s_{n} = 1 - \frac{ \sigma_{n}^{2} (\tilde{y}) }
{ \tau_{n}^{2} (\tilde{y}) },
\end{equation*}
were $\tau_{n} (\tilde{y})$ is the prior standard
deviation of $\tilde{\theta}_{n}$.  The posterior shrinkage
quantifies how much the posterior distribution contracts 
from the initial prior distribution.

An ideal experiment is extremely informative, with large 
shrinkage for every observation, while also being accurate, 
with small $z$-scores for every observation.  In this case
the distribution of posteriors derived from prior predictive
observations should concentrate towards small $z$-scores
and large posterior shrinkages for each parameter component.  
On the other hand, small posterior shrinkage indicates an 
experiment that poorly identifies the given parameter 
component, while large $z$-scores indicates inferences 
biased away from the true data generating process.  

\begin{figure}
\centering
\subfigure[] {\includegraphics[width=3.1in]{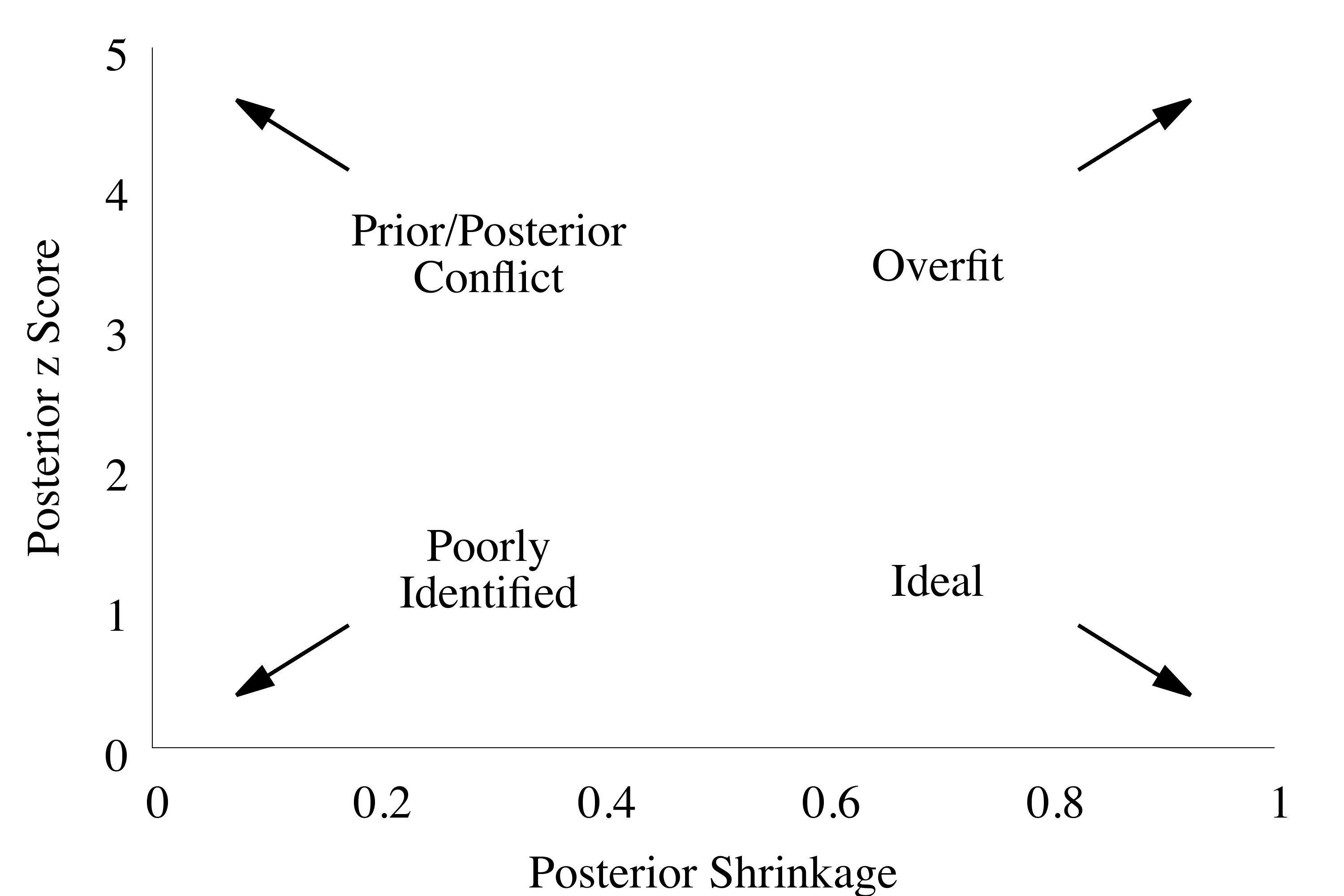} }
\subfigure[] {\includegraphics[width=2.9in]{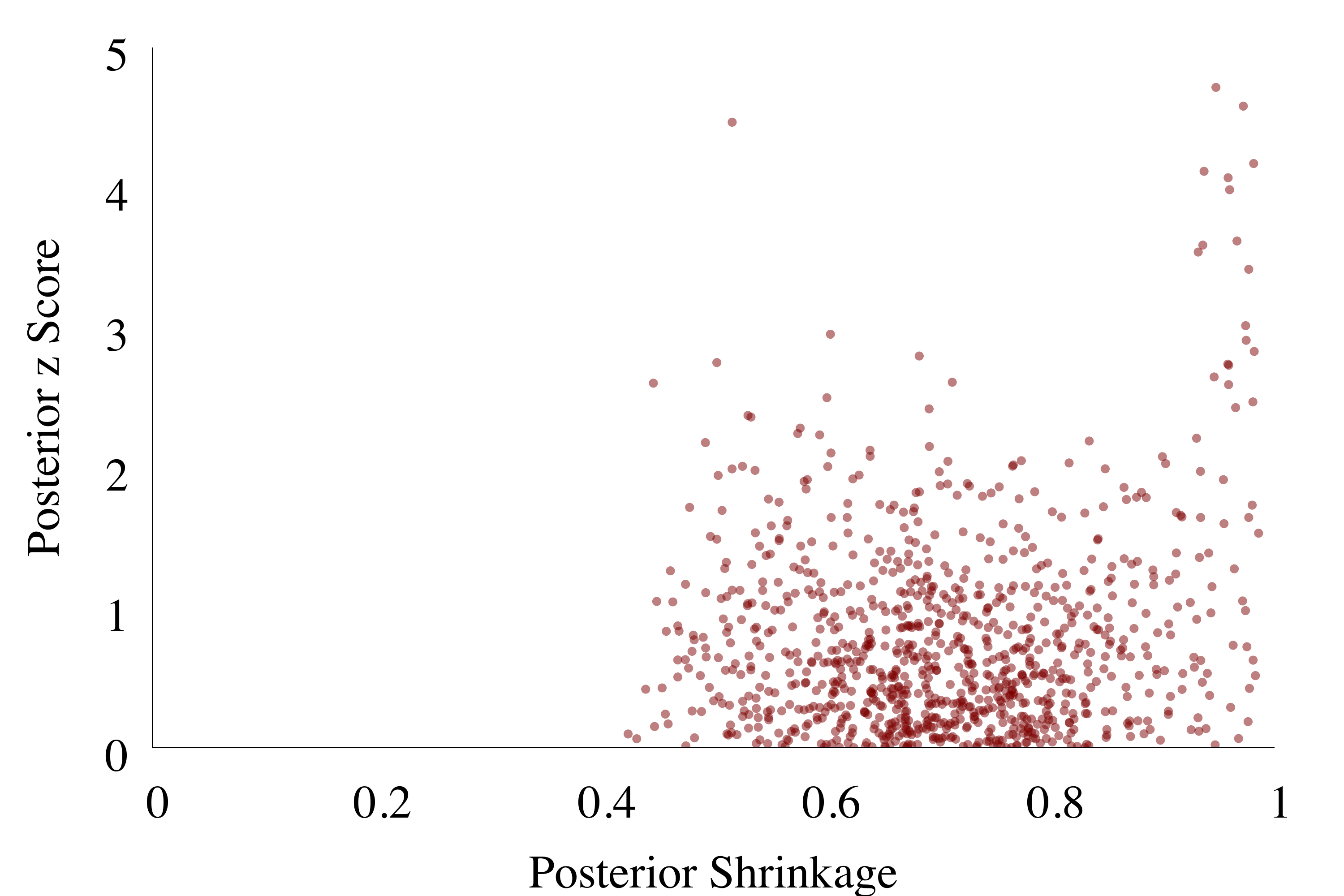} }
\subfigure[] {\includegraphics[width=2.9in]{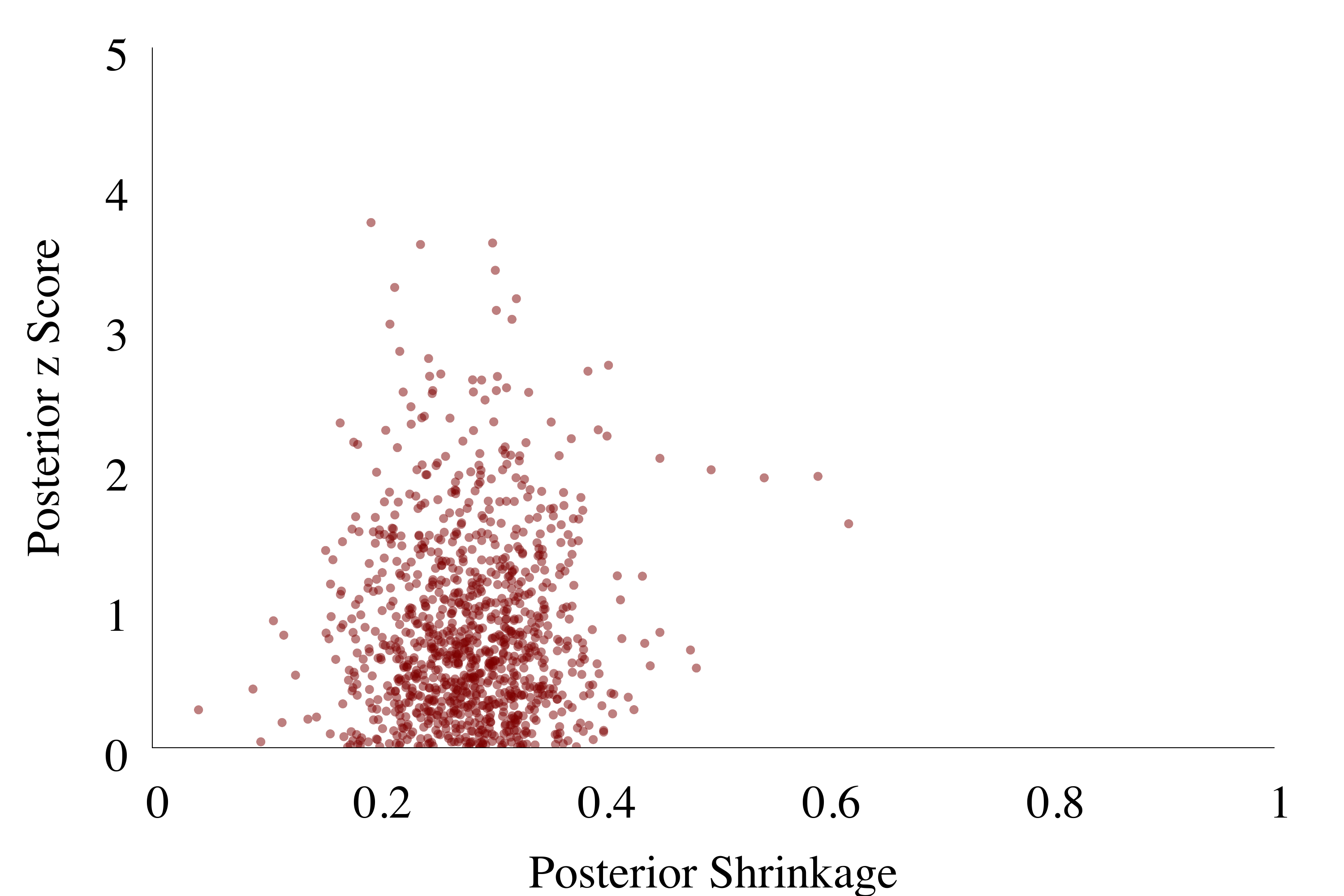} }
\caption{(a) The distribution of posterior $z$-scores and posterior 
shrinkages with respect to the Bayesian joint distribution identifies
well-known pathologies of Bayesian inference. The sensitivity of
the posterior to the observations within the scope of a given model
can be summarized by sampling from the Bayesian joint distribution,
constructing posteriors, and then plotting the corresponding
posterior $z$-scores and shrinkages.  (b) This model demonstrates 
good behavior for most observations, with a small tail of outcomes 
that overfit to a few observations. (c) This model is weakly identified, 
with the information introduced by the data doing little to improve 
upon the information encoded in the prior distribution.}
\label{fig:bayesian_eye_chart}
\end{figure}

We can readily visualize this behavior by plotting the 
posterior z- score verses the posterior shrinkage.
Concentration to the top right of this plot indicates 
overfitting, while concentration to the top left indicates 
a poorly-chosen prior that biases the model configuration 
space away from the presumed true data generating process 
(Figure \ref{fig:bayesian_eye_chart}a).  Because the 
Bayesian joint distribution considers only those true data 
generating consistent with the prior,  however, this latter 
behavior should be impossible within the scope of a 
model-based sensitivity analysis.

By investigating this simple summary we can quickly 
identify problems with our experimental design
(Figure \ref{fig:bayesian_eye_chart}b, c).  A scatter plot 
that combines the outcomes for all of the parameters components
into one plot first summarizes the aggregate performance of the 
entire model, and then individual plots for each parameter
component can be used to isolate the source of any noted
pathological behavior.

\subsection{Limitations of Model-Based Calibration}

The ultimate limitation of model-based calibration is its 
dependence on the model configuration space.  Any model-based
sensitivities or guarantees claimed by model-based calibrations 
rely on the model configuration space being rich enough to capture 
the true data generating process, or at least contain model 
configurations that approximate it sufficiently well.  

Unfortunately it is difficult to quantify how these guarantees
might change as the as the model configurations become worse
approximations to the true data generating process.  Consequently 
it is up to the user to verify the sufficiency of the assumed model 
configuration space with, for example, predictive validations such 
as residual analysis for frequentist point estimators or posterior 
predictive checks for Bayesian analyses.

Another point of fragility of model-based sensitivity and 
calibrations is that they apply only for the exact models and 
decisions being considered.  If those models or decisions are 
tweaked then the guarantees no longer need apply.  The only 
way to ensure valid calibrations is to recompute them every 
time the experiment is modified.  An consequence of this 
fragility is that any sensitivity or calibration is suspect 
whenever the construction of the model itself depends on the 
observed data!  The only rigorous way to maintain the validity 
of these results is to consider a larger model that incorporates 
this implicit dependence of the model configuration space on the
observed data.  

Because observations are often used to critique and ultimately 
tune the model, this vulnerability is almost impossible to avoid 
in practice.  Consequently model-based calibration is perhaps 
best considered as a tool for identifying poorly inferential
behaviors in a model rather than making absolute guarantees 
about its performance.

\section{Calibrating Discovery Claims}

One of the most common decisions made in the applied sciences
is whether or not to claim that a phenomenon in the system being 
studied exists or doesn't exist.  Whether such discovery claims
are good scientific practice is debatable, but given their
prevalence it is important to be able to calibrate these
decisions regardless.

Many decision making processes have been developed within in 
both statistics and applied fields, and many of these methods 
have come under recent scrutiny given their failure to replicate 
in subsequent experiments.  The underlying issue in these failed 
replications is often poor calibration of the original discovery 
claim.

In this section I review how discovery claims can be 
constructed from a statistical model both in the frequentist 
and Bayesian paradigms and discuss some of the practical 
issues with their calibration.

\subsection{Partitioning the Model Configuration Space}

In order to decide on the presence of a phenomenon we
need to partition the model configuration space into
those model configurations that are influenced by the
phenomenon and those that are not.  For example we might
partition the model configuration space into data
generating processes containing only background sources
and those containing both signal and background sources.
Alternatively we might consider a partition into model
configurations where two phenomena manifest distinct
behaviors and those where they behave identically.
Discovery claims are then informed by which of the two 
partitions is more consistent with the observed data.

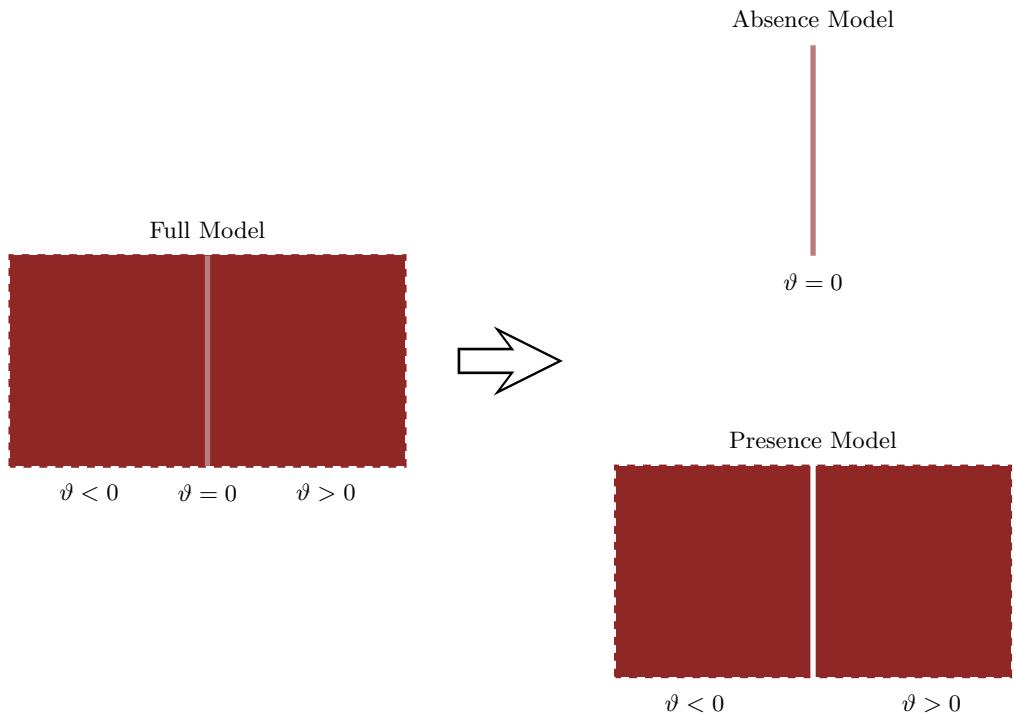
\begin{figure*}
\centering
\begin{tikzpicture}[scale=0.35, thick]
  % Full model
  \fill [fill=dark] (0, 0) rectangle (15, 8);
  \draw [color=dark, very thick, dashed] (0, 0) rectangle (15, 8);
  \node at (7.5, 9) {Full Model};
  
  \draw [color=mid, line width=2] (7.5, 0) -- (7.5, 8);
  
  \node at (3, -1) {$\vartheta < 0$};
  \node at (7.5, -1) {$\vartheta = 0$};
  \node at (12, -1) {$\vartheta > 0$};
  
  % Transition
  \draw[->, >=stealth, line width=10] (17, 4) -- (21, 4);
  \draw[->, >=stealth, line width=8, white] (17.1, 4) -- (20.8, 4);
  
  % Presence Model
  \fill [fill=dark] (23, -8) rectangle (38, 0);
  \draw [color=dark, very thick, dashed] (23, -8) rectangle (38, 0);
  \node at (30.5, 1) {Presence Model};
  
  \draw [color=white, line width=2] (30.5, -8.25) -- (30.5, 0.25);
  
  \node at (26, -9) {$\vartheta < 0$};
  \node at (35, -9) {$\vartheta > 0$};
  
  % Absence Model
  \draw [color=mid, line width=2] (30.5, 8) -- (30.5, 16);
  \node at (30.5, 17) {Absence Model};
  
  \node at (30.5, 7) {$\vartheta = 0$};

\end{tikzpicture}
\caption{When claiming a discovery we first identify the
\emph{phenomenological} parameters, $\vartheta$, which quantify
the behavior of a phenomena of interest.  The remaining
\emph{nuisance} parameters quantify the behavior of the 
environment in which the phenomenon interacts and the the 
measurement process itself.  When $\vartheta = 0$ identifies 
a vanishing phenomenon we can partition the full model 
configuration space into an \emph{absence model} of those 
model configurations satisfying $\vartheta = 0$ and a 
\emph{presence model} of those model configurations satisfying 
$\vartheta \ne 0$.  Discovery is claimed when the presence model 
is strongly preferred over the absence model by an observation, 
for various definitions of ``preferred''.
}
\label{fig:discovery}
\end{figure*}

Let the phenomenon of interest be characterized
with a subset of parameters, $\vartheta = \varpi(\theta)$, 
where $\varpi$ projects the total parameter space onto
the phenomenological parameters of interest.  Additionally 
assume that the parameters are structured such that 
$\vartheta = 0$ identifies those data generating processes 
not influenced by the phenomenon being considered. In this 
case the model configuration space partitions into an
\emph{absence model}
\begin{equation*}
\Theta_{1} = \left\{ \theta \in \Theta \mid 
\varpi(\theta) = 0 \right\},
\end{equation*}
and a complementary \emph{presence model},
\begin{equation*}
\Theta_{2} = \left\{ \theta \in \Theta \mid 
\varpi(\theta) \ne 0 \right\}
\end{equation*}
(Figure \ref{fig:discovery}).
This includes the case where $\vartheta$ is constrained
to be positive, in which case the presence model reduces
to $\theta > 0$, or the more general case where $\vartheta$
is unconstrained and the presence model includes all
positive and negative, but non-zero, parameters.

For example, consider a model where our observations
are generated from overlapping signal and background
sources, $\mu_{s}$ and $\mu_{b}$ respectively, with
Gaussian measurement variability $\sigma$.  This 
yields the model configuration space
\begin{equation*}
\pi(y \mid \mu_{s}, \mu_{b}, \sigma)
=
\mathcal{N} (y \mid \mu_{s} + \mu_{b}, \sigma).
\end{equation*}
If we are interested in understanding the signal then 
we would consider the projection
\begin{equation*}
\mu_{s} = \vartheta = \varpi(\mu_{s}, \mu_{b}, \sigma),
\end{equation*}
with the absence model defined as $\mu_{s} = 0$,
regardless of the value of the nuisance parameters
$\mu_{b}$ and $\sigma$, and the presence model as 
the complement with $\mu_{s} \neq 0$.

Because we don't know which of the two partitions
contains the true data generating process we have to 
calibrate our decisions with respect to \emph{both}.
In particular we have to consider the probability
of claiming a discovery when the true data generating
process is in the presence model and when it is in
the absence model.  We can fully characterize the
four possible outcomes with two probabilities: the
\emph{false discovery rate}, $\mathrm{FDR}$ and 
the \emph{true discovery rate}, $\mathrm{TDR}$
(Table \ref{tab:two_way_discovery}).  In classical
statistics the false discovery rate is also known
as the \emph{Type I error} while one minus the
true discovery rate is also known as the \emph{Type
II error}.

\begin{table}
  \centering
  \renewcommand{\arraystretch}{1.5}
  \begin{tabular}{ccc}
    \rowcolor[gray]{0.9} 
    \textbf{Truth} & \multicolumn{2}{c}{\textbf{Decision}} \\
    \rowcolor[gray]{0.9} 
    & Claim $\theta^{*} \in \Theta_{1}$  
    & Claim $\theta^{*} \in \Theta_{2}$ \\
    $\theta^{*} \in \Theta_{1}$ (No Phenomenon) 
    & $\mathrm{FDR}$ & $1 - \mathrm{FDR}$   \\
    $\theta^{*} \in \Theta_{2}$ (Phenomenon) 
    & $1 - \mathrm{TDR}$ & $\mathrm{TDR}$ \\
    & & 
  \end{tabular}
\caption{When preparing to claim a discovery we have to 
consider the range of inferential outcomes both for when
the true data generating process, $\theta^{*}$, is an 
element of the absence model, $\theta^{*} \in \Theta_{1}$,
and when it is an element of the presence model, 
$\theta^{*} \in \Theta_{2}$.  Given the false discovery rate, 
$\mathrm{FDR}$, and true discovery rate, $\mathrm{TDR}$, we 
can compute any expected loss with respect to the claiming
a discovery or not.}
\label{tab:two_way_discovery}
\end{table}

Given these probabilities we can compute the expected loss 
once we have assigned losses to the possible decision outcomes.  
For example, let $L_{1}$ be the loss associated with claiming 
a discovery when the true data generating process is in the 
absence model and $L_{2}$ the possibly negative loss associated 
with claiming a discovery when the true data generating process 
is in the presence model.  The expected loss for claiming a 
discovery is then given by
\begin{equation*}
L_{\text{claim discovery}} 
= 
(1 - \mathrm{FDR}) \, L_{1} + \mathrm{TDR} \, L_{2}.
\end{equation*}

In practice we need not limit ourselves to dichotomous 
decisions.  We could also consider a decision process
that claims the phenomenon exists, claims the phenomenon
doesn't exist, or makes no claim at all.  This process
would be characterized by six probabilities, four of 
which are independent (Table \ref{tab:three_way_discovery}).  
Importantly the expected loss of a discovery claim can be 
miscalculated if we ignore the possibility that an analysis 
may not be reported for some observations and the two
additional degrees of freedom needed to quantify the
expected loss in this more general circumstance.

\begin{table}
  \centering
  \renewcommand{\arraystretch}{1.5}
  \begin{tabular}{cccc}
    \rowcolor[gray]{0.9} 
    \textbf{Truth} & \multicolumn{3}{c}{\textbf{Decision}} \\
    \rowcolor[gray]{0.9} 
    & Claim $\theta^{*} \in \Theta_{1}$  
    & Claim $\theta^{*} \in \Theta_{2}$
    & Claim Nothing \\
    $\theta^{*} \in \Theta_{1}$ (No Phenomenon) & $p_1$ & $p_2$ & $p_3$   \\
    $\theta^{*} \in \Theta_{2}$ (Phenomenon)  & $p'_1$ & $p'_2$ & $p'_3$ \\
    & & &
  \end{tabular}
\caption{A decision process where we can avoid making any
claim about the phenomenon of interest by not publishing
our analysis is characterized by six probabilities, four 
of which are independent because of the normalization 
constraints, $\sum_{n = 1}^{3} p_{n} = 1$ and 
$\sum_{n = 1}^{3} p'_{n} = 1$.
}
\label{tab:three_way_discovery}
\end{table}

\subsection{Frequentist Null Hypothesis Significance Testing}

The conventional approach to claiming discoveries
in a frequentist framework is the \emph{null hypothesis
significance testing} framework.  Here the \emph{null
hypothesis} that the true data generating process falls 
into the absence model is treated as something of a 
strawman set up to be rejected by observation.  
In order to reject the null hypothesis we consider
how extreme an observation is with respect to the
model configurations in the absence model. The more 
extreme our rejection threshold the smaller the false 
discovery rate should be.

Naively, if the null hypothesis is rejected then we 
are left with only the \emph{alternative hypothesis} 
that the true data generating process falls into the 
presence model.  That said, we can't simply reject the 
null hypothesis in isolation --- a poor fit to the null
hypothesis does \emph{not} imply that the alternative 
hypothesis is any more consistent with the observation!  
At the very least we have to consider also how likely 
we are to reject the null hypothesis when the alternative 
hypothesis is true.

Exactly how the null hypothesis significance testing
framework is implemented depends on the structure of
the null and alternative hypotheses.  The procedure
is straightforward for simple hypotheses but quickly
becomes difficult to implement in practice as the
hypotheses become more complex.

\subsubsection{Point Hypotheses}

The simplest case of null hypothesis significance
testing is when both the null hypothesis and 
alternative hypothesis are \emph{point hypotheses}
consisting of a single model configuration each.
In this case we'll denote the lone data generating 
processes in the absence model $\pi_{N} (y)$ and 
the lone data generating process in the precence
model $\pi_{A} (y)$.

For a point null hypothesis we can quantify the 
extremity of an observation, $\tilde{y}$, with a
tail probability or \emph{$p$-value},
\begin{equation*}
p (\tilde{y} ) = 
\int_{\tilde{y}}^{\infty} \mathrm{d} y \,
\pi_{N} (y).
\end{equation*}
The integral might be computed analytically or with 
numerical methods such as quadrature for 
low-dimensional measurement spaces and Monte Carlo
for high-dimensional measurement spaces.

If we reject the null hypothesis when
\begin{equation*}
p (\tilde{y} ) < 1 - \alpha
\end{equation*}
for some \emph{significance}, $\alpha$, then by
construction the false discovery rate of our
claim will be
\begin{align*}
\mathrm{FDR} 
&= 
\int_{Y} \mathrm{d} y \, \pi_{N} (y) \,
\mathbb{I} \left[ \, p ( y ) < 1 - \alpha \right]
\\
&=
\int_{0}^{1} \mathrm{d} \tilde{p} \;
 \mathbb{I} \left[ \, \tilde{p} < 1 - \alpha \right]
\\
&=
1 - \alpha.
\end{align*}
By tuning the significance of the null hypothesis test
we can immediately achieve whatever false discovery rate 
is desired in a given application.

The true discovery rate, also known as the $\emph{power}$ 
of the null hypothesis test, is the average null $p$-value 
with respect to the alternative data generating process,
\begin{equation*}
\mathrm{TDR} = \beta = 
1 - \int_{Y} \mathrm{d} y \, \pi_{A} (y) \, p (y).
\end{equation*}
Provided that the power is sufficiently high, observations 
for which the null hypothesis is rejected will be more 
consistent with the alternative hypothesis \emph{in 
expectation}.  There is no guarantee, however, that the 
alternative will actually be more consistent for every
observation.

Unlike the false discovery rate, the true discovery rate 
is a consequence of the assumed model and cannot be tuned
once a significance has been set.  Consequently 
unsatisfactorily low true discovery rates can be
remedied only by modifying the experimental circumstances,
for example by increasing the number of observations
included in each measurement.

\subsubsection{Point Null Hypotheses and Complex Alternative Hypotheses}
\label{point_null_complex_alt}

Null hypothesis significance testing becomes more
complicated when the model configuration space no
longer consists of just two data generating processes
and both hypotheses cannot be point hypotheses.
Consider next the situation where the null hypothesis
is still given by a single data generating process,
$\theta = 0$, and the alternative hypothesis contains
the remaining model configurations $\theta \neq 0$.
This might arise, for example, when our model contains
only one phenomenological parameter and $\theta = 0$
uniquely defines the circumstance where the phenomenon 
is absent.

As before we can define the $p$-value,
\begin{equation*}
p (\tilde{y} ) = 
\int_{\tilde{y}}^{\infty} \mathrm{d} y \,
\pi (y \mid \theta = 0),
\end{equation*}
and then reject the null hypothesis when
\begin{equation*}
p (\tilde{y} ) < 1 - \alpha
\end{equation*}
to ensure a given false discovery rate.

\begin{figure}
\centering
\subfigure[] {\includegraphics[width=2.9in]{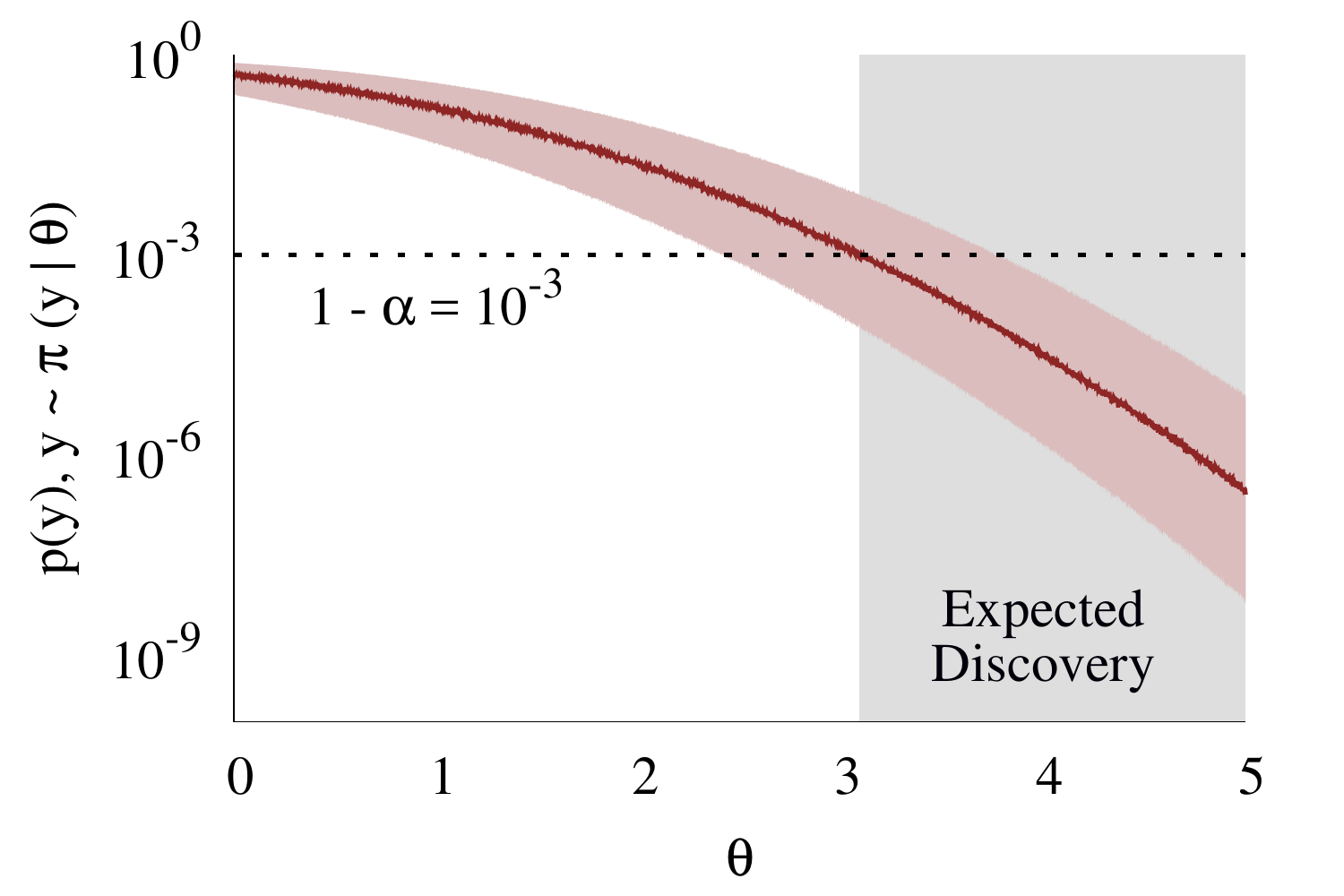} }
\subfigure[] {\includegraphics[width=2.9in]{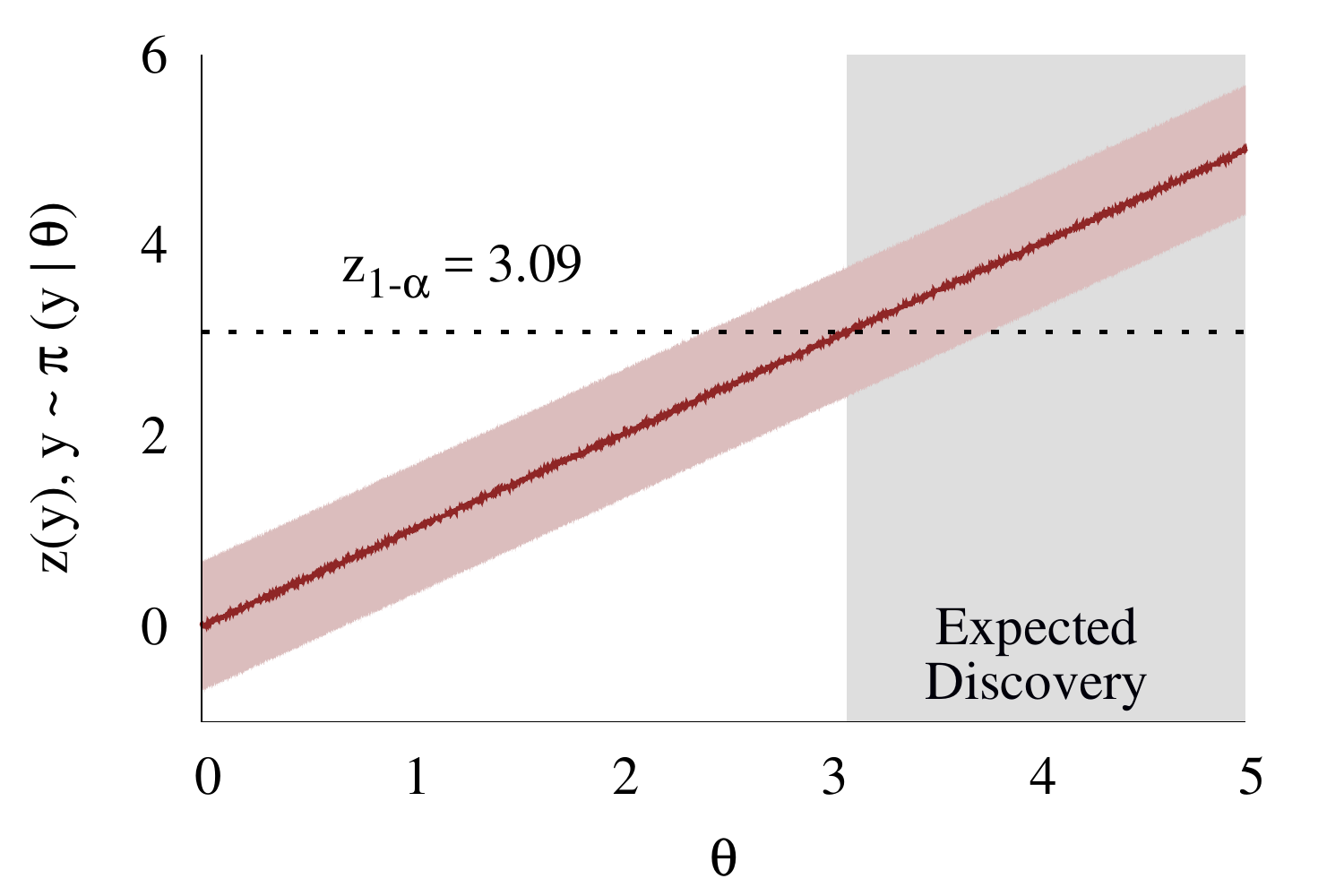} }
\caption{When the model configuration space is one-dimensional
we can visualize the sensitivity of the null $p$-values for each
of the model configurations in the alternative hypothesis, 
$\theta \neq 0$.  This allows us to quickly communicate the 
distribution of (a) the null $p$-values, $p(y)$ or the equivalent
(b) standard scores $z(y) = \Phi^{-1}(1  - p(y))$. Here the
distributions are summarized with their medians in dark red and
with the quartiles in light red.  In particular we can identify 
which $\theta$ in the alternative hypothesis achieve a given true 
discovery rate, here $0.999$, by seeing where the central value 
of the distributions surpass a dashed line.}
\label{fig:freq_disc_sens}
\end{figure}

\begin{figure}
\centering
\includegraphics[width=4in]{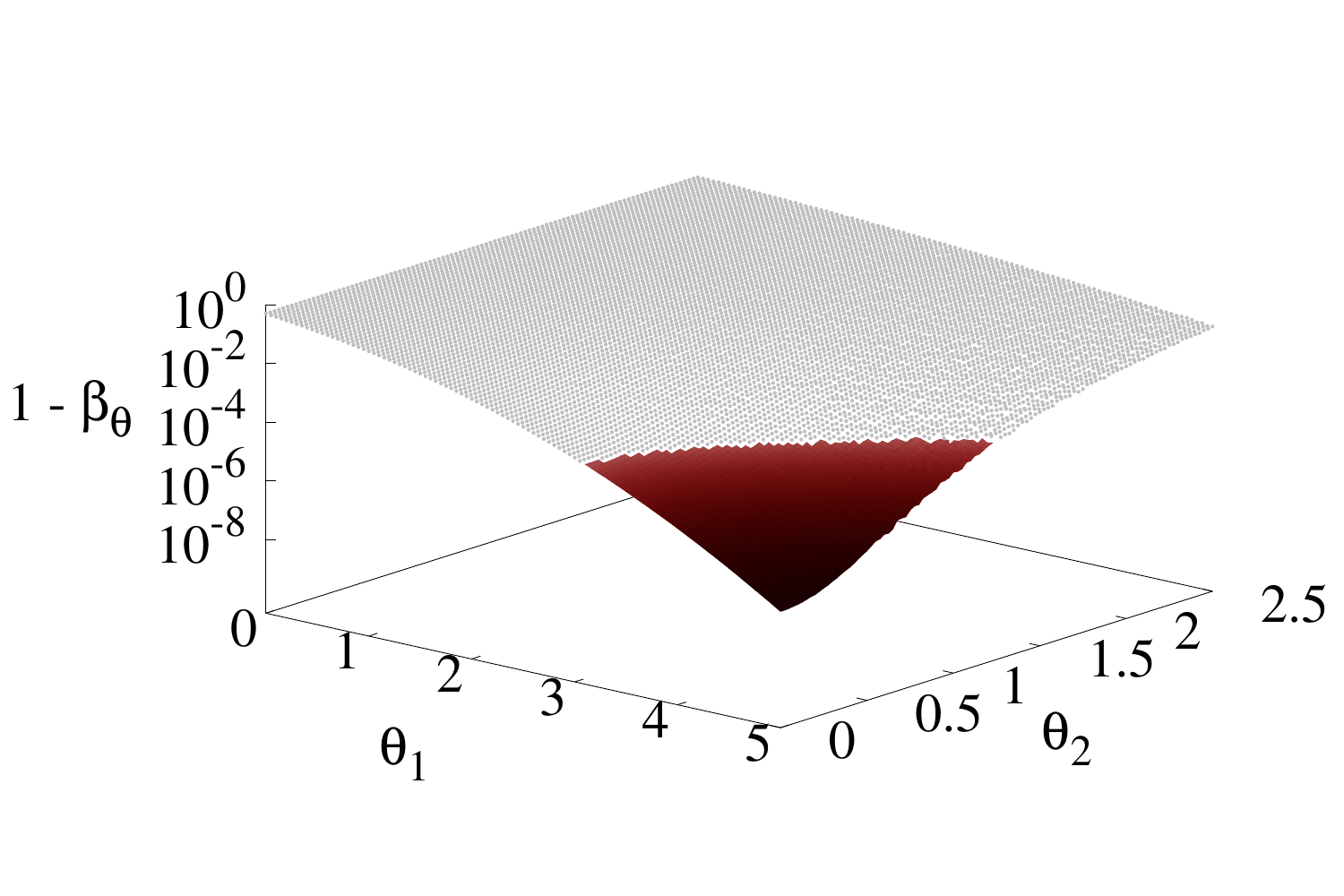}
\caption{When the model configuration space is two-dimensional
we can visualize how the power, $\beta_{\theta}$, varies with
the model configurations in the alternative hypothesis, here
$(\theta_{1}, \theta_{2}) \neq (0, 1)$.  As in the one-dimensional
case we can vary the decoration to identify those model 
configurations achieving a given true discovery rate, here $0.999$, 
shown in red.}
\label{fig:freq_disc_sens_2d}
\end{figure}

Unfortunately there is no longer a unique way of 
defining a power that gives the true discovery rate
because the true discovery rate will, in general,
be different for each of the model configuration
in the alternative hypothesis,
\begin{equation*}
\beta_{\theta} = 
1 - \int_{Y} \mathrm{d} y \, \pi (y \mid \theta) \, p (y),
\, \theta \neq 0.
\end{equation*}
If $\Theta$ is one-dimensional then we can visualize
the sensitivity of the $p$-values as a function of 
$\theta$ (Figure \ref{fig:freq_disc_sens}).  When 
$\Theta$ is two-dimensional we can no longer visualize 
the full variation of the sensitivity distribution, but 
we can visualize the variation of a summary statistic 
such as the power (Figure \ref{fig:freq_disc_sens_2d}).  
Visualizations allow us to identify regions in the 
alternative model configuration space of high power, 
but they do not define a unique power or true discovery 
rate for the test.

One immediate strategy is to define an overall power is to
consider the minimum power over all of the model configurations 
in the alternative hypothesis,
\begin{equation*}
\beta \equiv \min_{\theta \neq 0} \beta_{\theta}.
\end{equation*}
Provided that we could accurately compute the minimum, this
definition would ensure that the overall power lower bounds
the true discovery rate for all model configurations in
the alternative hypothesis.  Unfortunately, when the null 
hypothesis is \emph{nested} within the alternative hypothesis
the power can become arbitrarily small for the alternative
model configurations in the neighborhood around the lone
null model configuration.  Consequently in practice we will
generally be able to claim calibrated discoveries \emph{only 
for a subset of the data generating processes in the alternative 
hypothesis}.

In order to visualize how the power varies with the
phenomenological parameters, $\vartheta$, we might also
consider defining \emph{conditional powers}.  If the 
parameters partition into $(\vartheta, \sigma)$, with 
$\vartheta \in \varpi(\Theta)$ and
$\sigma \in \varpi(\Theta)^{C}$, then
we could define the conditional power as
\begin{equation*}
\beta_{\vartheta} = 
1 - \max_{\sigma \in \varpi(\Theta)^{C}} 
\int_{Y} \mathrm{d} y \, \pi (y \mid \vartheta, \sigma) \, p (y).
\end{equation*}
Because the optimization of the nuisance parameters,
$\sigma$, may be infeasible, profile likelihood
methods are often utilized to approximate the conditional 
powers for visualization.  As with any application of 
asymptotics, the validity of this approximations must be 
carefully verified for the visualization to be useful.

Finally we might acknowledge the conceptual advantage
of having dual point hypotheses and consider not one null 
hypothesis test but rather \emph{an infinite number of tests},
where each tests is defined with respect to the point null 
hypothesis $\theta = 0$, and one of the alternative model
configurations, $\theta' \neq 0$.  When we reject the null
hypothesis we reject it for \emph{any} of the alternatives.
The preponderance of alternative hypotheses, however,
significantly increases the false discovery rate unless
we apply a \emph{multiple comparison correction} to
the significance threshold.  This increase in the false 
discovery rate by separating the alternative hypothesis 
into many point hypothesis is also known as the ``look 
elsewhere effect'' in particle physics.

\subsubsection{Complex Hypotheses}

Unsurprisingly, implementing null hypothesis significance
testing becomes all the more difficult when neither the
null hypothesis nor the alternative hypothesis are point
hypotheses.  Given the influence of systematic and 
environmental factors present in any experiment we rarely 
if ever enjoy point hypotheses when using realistic models.

When there are multiple model configurations in the null 
hypothesis we have to consider them all.  For example we
might define the $p$-value to be the smallest tail probability
across all of the null model configurations,
\begin{equation*}
p (\tilde{y} ) = 
\min_{\theta \in \Theta_{1}} 
\int_{\tilde{y}}^{\infty} \mathrm{d} y \,
\pi (y \mid \theta).
\end{equation*}
If we then reject the null when this minimal $p$-value is 
less than $1 - \alpha$ then the false discovery rate will 
be at least $\alpha$ for every data generating process in 
the null hypothesis.

Power and true discover rate calculations proceed as above,
with all of the potential complications.

In this general case the computation of the optima needed
to bound the false and true discovery rates becomes a 
particularly significant computational burden that must
be addressed with a careful combination of principled
assumptions and approximations.

\subsubsection{The Likelihood Ratio Test}

One of the difficulties with the null hypothesis significance
testing framework presented so far is the need to compute
tail probabilities over the measurement space.  When the
measurement space is more than a few dimensions these tail
probabilities are difficult to accurately approximate even
with substantial computational resources available.  A
better strategy is to construct a lower-dimensional summary
of the measurement space that captures the differences between
the null and alternative hypotheses while admitting tests 
that are easier to implement.

Perhaps the most ubiquitous summary for testing is the
\emph{likelihood ratio}
\begin{equation*}
\lambda(\tilde{y}) = 
\frac{ \max_{\theta \in \Theta_{1}} \pi(\tilde{y} \mid \theta) }
{ \max_{\theta \in \Theta} \,\, \pi(\tilde{y} \mid \theta) },
\end{equation*}
which admits the \emph{likelihood ratio test} where we
reject the null hypothesis if $\lambda(\tilde{y}) < \lambda_{0}$ 
for some $0 < \lambda_{0} < 1$.

The false and true discovery rates of the likelihood ratio
test intimately depend on the threshold, $\lambda_{0}$, and
the particular structure of the model configuration space.
Consequently without further assumptions the likelihood ratio 
test has to be explicitly calibrated for every application.

The assumption of asymptotics, however, admits an analytic
calibration of the likelihood ratio test.  \emph{Wilk's
Theorem} demonstrates that, under the typical asymptotic
conditions, the distribution of the logarithm of the likelihood 
ratio with respect to the model configurations in the null
hypothesis asymptotically approaches a $\chi^{2}$ distribution,
\begin{equation*}
- 2 \log \lambda \sim \chi^{2}_{K},
\end{equation*}
with degrees of freedom,
\begin{equation*}
K = \mathrm{dim}(\Theta) - \mathrm{dim}(\Theta_{1}).
\end{equation*}
Consequently the false discovery rate for a given 
threshold, $-2 \log \lambda_{0}$, can be calculated
by looking up the corresponding $\chi^{2}_{K}$ tail
probability.

Indeed theoretical analysis shows that the likelihood
ratio test is the optimal test in this asymptotic
regime.  Many popular tests that have been developed
in applied fields, such as the Feldman-Cousins test
\citep{FeldmanEtAl:1998}, are actually instances of the 
likelihood ratio test for specific classes of models.

\subsection{Bayesian Model Comparison}

Bayesian model comparison is an immediate consequence
of extending a probabilistic treatment to the absence and
presence partitions of the model configuration space.  
Given that it's not conceptually more difficult, however, 
let's consider the more general case where we are interested 
in selecting between one of $N$ models,
$\left\{\mathcal{M}_1, \ldots, \mathcal{M}_n\right\}$.  

Each model configuration space can have different dimensions,
but integrating the parameters out of the corresponding
Bayesian joint distribution gives a \emph{marginal likelihood}
over the common measurement space,
\begin{align*}
\pi( y \mid \mathcal{M}_n) 
&=
\int_{\Theta} \mathrm{d} \theta \,
\pi( y, \theta \mid \mathcal{M}_n)
\\
&=
\int_{\Theta} \mathrm{d} \theta \,
\pi(y \mid \theta, \mathcal{M}_n) \,
\pi(\theta \mid \mathcal{M}_n)
\\
&=
\mathbb{E}_{\pi(\theta \mid \mathcal{M}_n)} 
[ \pi(y \mid \theta, \mathcal{M}_n) ].
\end{align*}
The marginal likelihood is also often known as the 
\emph{Bayes factor} or \emph{evidence} in some fields.

Given the marginal likelihoods we can construct a joint
distribution over the measurement and model spaces,
\begin{equation*}
\pi(y, \mathcal{M}_n) = \pi(y \mid \mathcal{M}_n) \, \pi(\mathcal{M}_n),
\end{equation*}
from which Bayes' Theorem gives the \emph{model posteriors},
\begin{equation*}
\pi(\mathcal{M}_n \mid \tilde{y} )
=
\frac{ \pi(\tilde{y}, \mathcal{M}_n) }{ \pi(\tilde{y}) }
=
\frac{ 
\pi(\tilde{y} \mid \mathcal{M}_n) \, 
\pi(\mathcal{M}_n) }
{ 
\sum_{n'=1}^{N} \pi(\tilde{y} \mid \mathcal{M}_n') \, 
\pi(\mathcal{M}_n') }.
\end{equation*}

In particular, given only two models, $\mathcal{M}_{1}$ and
$\mathcal{M}_{2}$, we are immediately guided to select the first 
when it exhibits a higher model posterior density,
\begin{align*}
\pi(\mathcal{M}_1 \mid \tilde{y} ) &> \pi(\mathcal{M}_2 \mid \tilde{y} )
\\
\frac{ 
\pi(\tilde{y} \mid \mathcal{M}_1) \, 
\pi(\mathcal{M}_1) }
{ 
\sum_{n'=1}^{N} \pi(\tilde{y} \mid \mathcal{M}_n') \, 
\pi(\mathcal{M}_n') }
&>
\frac{ 
\pi(\tilde{y} \mid \mathcal{M}_2) \, 
\pi(\mathcal{M}_2) }
{ 
\sum_{n'=1}^{N} \pi(\tilde{y} \mid \mathcal{M}_n') \, 
\pi(\mathcal{M}_n') }
\\
\pi(\tilde{y} \mid \mathcal{M}_1) \, \pi(\mathcal{M}_1)
&>
\pi(\tilde{y} \mid \mathcal{M}_2) \, \pi(\mathcal{M}_2)
\\
\frac{ \pi(\tilde{y} \mid \mathcal{M}_1) }
{ \pi(\tilde{y} \mid \mathcal{M}_2) }
&>
\frac{ \pi(\mathcal{M}_2) }{  \pi(\mathcal{M}_1) }.
\end{align*}
In words, we select $\mathcal{M}_{1}$ when the \emph{odds ratio},
$\pi(\tilde{y} \mid \mathcal{M}_1) \, / 
\, \pi(\tilde{y} \mid \mathcal{M}_2) $,
surpasses a threshold defined by the the relative prior
probabilities of the two models, 
$ \pi(\mathcal{M}_2) /  \pi(\mathcal{M}_1)$.  Interestingly 
this procedure resembles the likelihood ratio test where we 
use marginal likelihoods instead of maximum likelihoods and 
the testing threshold is constructed from our prior 
distributions.

Calibration of this \emph{Bayesian model selection} then
proceeds as with the calibration of any Bayesian inference 
or decision making processs.
\begin{enumerate}
\item We first sample a true model from the model prior,
\begin{equation*}
\mathcal{M}_{\tilde{n}} \sim \pi(\mathcal{M}_{n}).
\end{equation*}
\item Then we sample a true model configuration from 
the subsequent prior distribution,
\begin{equation*}
\tilde{\theta} \sim \pi(\theta \mid \mathcal{M}_{\tilde{n}}).
\end{equation*}
\item Next we sample an observation from that model 
configuration,
\begin{equation*}
\tilde{y} \sim \pi(y \mid \tilde{\theta}, \mathcal{M}_{\tilde{n}}),
\end{equation*}
\item Finally we calculate the marginal likelihoods
to inform model selection,
\begin{equation*}
\pi (\tilde{y} \mid \mathcal{M}_{n}),
\end{equation*}
and estimate the corresponding discovery rates for each
model,
\begin{equation*}
R_{\mathcal{M}_{n} \mid \mathcal{M}_{n'}}
\approx
\frac{ \sharp [ \text{ Select }{\mathcal{M}_{n}}
\text{ given observation from }\mathcal{M}_{n'} ] }
{ \sharp [ \text{ Observations from }\mathcal{M}_{n'} ] }.
\end{equation*}
\end{enumerate}

In particular, if we define $\mathcal{M}_{A}$ as the 
absence model and $\mathcal{M}_{P}$ as the presence 
model then the false discovery rate is estimated as
\begin{equation*}
\mathrm{FDR}
\approx
\frac{ \sharp [ \text{ Select }{\mathcal{M}_{P}}
\text{ given observation from }\mathcal{M}_{A} ] }
{ \sharp [ \text{ Observations from }\mathcal{M}_{A} ]}.
\end{equation*}
with the true discovery rate estimated as
\begin{equation*}
\mathrm{TDR}
\approx
\frac{ \sharp [ \text{ Select }{\mathcal{M}_{P}}
\text{ given observation from }\mathcal{M}_{P} ] }
{ \sharp [ \text{ Observations from }\mathcal{M}_{P} ] }.
\end{equation*}
In cases like this where there are only a few models being
considered it may also be easier to condition on each model
and compute the corresponding discovery rates one at a time
instead of sampling a model at each iteration.

Marginal likelihoods and Bayesian model selection arise 
immediately once we consider probabilities over the set of 
models.  Unfortunately the theoretical elegance of this 
approach does not always translate into practical utility.

First and foremost the marginal likelihood is extremely 
challenging to estimate, even in relatively simple problems.  
The structure of the integral frustrates typical computational 
tools like Markov chain Monte Carlo and necessitates more 
complex, and less well established, tools like nested sampling 
and simulated tempering.  Unfortunately these methods are 
poorly understood relative to the more established tools 
and consequently their implementations are still limited 
by our modest understanding.  In particular, quantification
of the accuracy of these methods is typically limited to 
only heuristics.

Beyond the computational issues, however, is a more subtle
conceptual issue.  The marginal likelihood evaluates a model
by comparing a given observation to \emph{all} of the model 
configurations in the model configuration space, each 
weighted by only the prior distribution.  Consequently even 
the smallest details of the prior distribution can 
significantly affect the marginal likelihood.

This is in stark contrast to the effect of the prior 
distribution on the posterior distribution.  Here the 
likelihood reduces the influence of model configurations 
inconsistent with an observation, obscuring much of the 
structure of the prior distribution.  Even seemingly 
irrelevant details of the prior distribution will still
strongly affect the marginal likelihoods, and the 
practice of constructing prior distributions to ensure 
only well-behaved posteriors is grossly insufficient for 
ensuring meaningful marginal likelihoods 
\citep{GelmanEtAl:2017}.

In practice the sensitivity of the marginal likelihoods,
and hence Bayesian model selection, to the intricate
details of the prior distribution manifests in strong 
dependencies on the observation and a fragility in the
corresponding model selection.  Small changes in the 
observation can cause significant changes in the marginal 
likelihoods, with the decision making process rapidly
vacillating amongst the possible models.  Fortunately
this behavior will manifest in sensitivity analyses
and poor false discovery rates and true discovery rates
and so it can be quantified provided that the test is
calibrated!

\subsection{Posterior Probability of The Region of Practical Equivalence}

One of the implicit difficulties in informing discovery 
claims as presented so far is that the absence model is 
singular with respect to the full model configuration 
space -- the absence model configuration space and the 
presence model configuration spaces are of different 
dimensionality.  Because of this the posterior probability 
for all of the model configurations in the absence will 
always be zero for a prior that is continuous across the 
full model configuration space.  The only way to admit 
non-zero posterior probabilities over both models is to 
assign infinitely more prior probability to those model 
configurations in the absence model relative to those in 
the presence model.

Bayesian model comparison avoids this issue by comparing
only marginal likelihoods and avoiding the individual
model posteriors altogether.   We can inform a discovery 
claim using only the posterior over the full model 
configuration space, however, if we absorb some of the
presence model into the absence model.  In particular, 
those model configurations in the presence model close 
to those in the absence model will generate nearly 
identical observations and hence indistinguishable 
inferences; an infinitesimally weak phenomenon will 
be impossible to differentiate from no phenomenon without 
an impractical amount of data.

This suggests that we redefine our absence model as
\begin{equation*}
\Theta_{1} = \left\{ \theta \in \Theta \mid 
\left| \vartheta(\theta) \right| \le \vartheta_{0} \right\},
\end{equation*}
with the presence model becoming
\begin{equation*}
\Theta_{2} = \left\{ \theta \in \Theta \mid 
\left| \vartheta(\theta) \right| > \vartheta_{0} \right\},
\end{equation*}
for some threshold $\vartheta_{0}$.  The neighborhood
around the absence model configurations, 
$\left| \vartheta(\theta) \right| \le \vartheta_{0}$, 
is known as the \emph{region of practical equivalence} 
\citep{Kruschke:2014}.  Notice that separating model 
configurations close to $\vartheta = 0$ from the presence 
model is not entirely dissimilar in what we had to do when 
considering the power of a complex alternative model in 
null hypothesis significance testing.

With this modification of the absence model we can
then claim a discovery when the posterior probability
in the region of practical equivalence is below a
given threshold,
\begin{equation*}
\int_{-\vartheta_{0}}^{\vartheta_{0}} \mathrm{d} \theta \,
\pi(\vartheta \mid \tilde{y} ) < 1 - \alpha,
\end{equation*}
where $\pi(\vartheta \mid \tilde{y} )$ is the
marginal posterior over the phenomenological parameters.

I have defined the formal decision making process here to 
superficially resemble that used in null hypothesis 
significance testing, but we could just as easily use
the complementary situation where the posterior probability
outside the region of practical equivalence is above
the given threshold,
\begin{equation*}
\int_{-\infty}^{-\vartheta_{0}} \mathrm{d} \theta \,
\pi(\vartheta \mid \tilde{y} )
+
\int_{\vartheta_{0}}^{\infty} \mathrm{d} \theta \,
\pi(\vartheta \mid \tilde{y} )
 > 1 - \alpha.
\end{equation*}

Calibration of this method, in particular the estimation
of the false discovery rate and true discovery rate,
immediately follows from the Bayesian calibration
paradigm.
\begin{enumerate}
\item We first sample a true model configuration from 
the prior distribution over the model configuration space,
\begin{equation*}
\tilde{\theta} \sim \pi(\theta).
\end{equation*}
\item Next we sample an observation from that model 
configuration,
\begin{equation*}
\tilde{y} \sim \pi(y \mid \tilde{\theta}),
\end{equation*}
\item Finally we reconstruct the posterior probability 
of the absence model,%
\begin{equation*}
P = \int_{-\vartheta_{0}}^{\vartheta_{0}} \mathrm{d} \vartheta \,
\pi(\vartheta \mid \tilde{y} ).
\end{equation*}
The false discovery rate follows as
\begin{equation*}
\mathrm{FDR}
\approx
\frac{ \sharp [ P \le 1 - \alpha\text{ and }
|\varpi(\tilde{\theta}) | < \vartheta_{0} ]}
{ \sharp [ | \varpi(\tilde{\theta}) | < \vartheta_{0} ] },
\end{equation*}
with the true discovery rate,
\begin{equation*}
\mathrm{TDR}
\approx
\frac{ \sharp [ P \le 1 - \alpha\text{ and }
|\varpi(\tilde{\theta}) | > \vartheta_{0} }
{ \sharp [ |\varpi(\tilde{\theta}) | > \vartheta_{0} ] }.
\end{equation*}
\end{enumerate}

\begin{figure}
\centering
\begin{tikzpicture}[scale=1, thick]
  \node at (0, 0) {\includegraphics[width=2.9in]{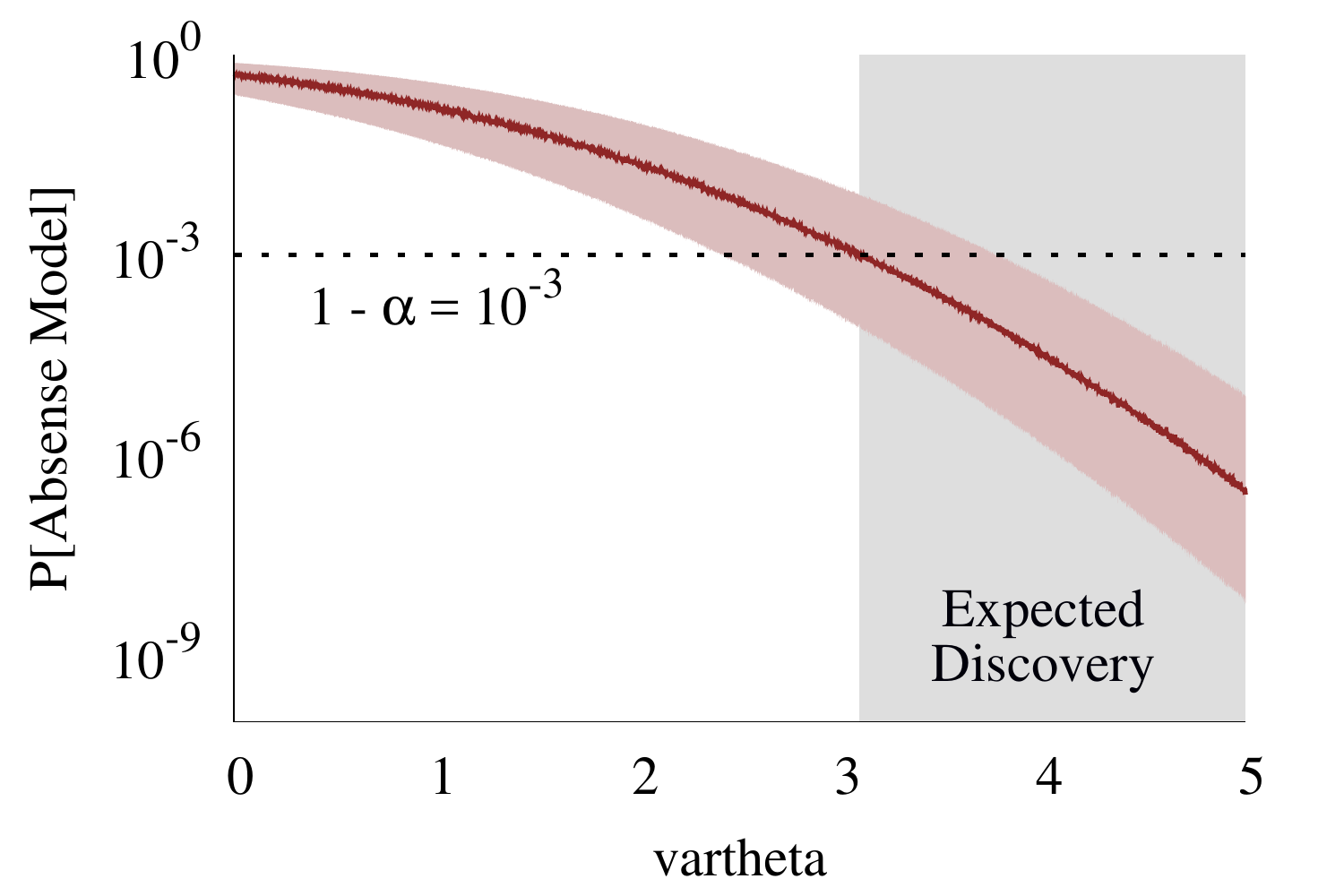}};
  \fill [color=white, text=black] (-1, -2.5) rectangle (2, -2)
  node[midway, align=center] { $\vartheta$ };
\end{tikzpicture}
\caption{By sampling from a conditional joint
distribution, $\pi(\sigma \mid \vartheta)$, we can 
visualize the sensitivity of the distribution of 
absence model posterior probabilities  with respect 
to a phenomenological parameter, $\vartheta$.  
The average of this distribution, shown here in dark
red, gives the conditional false discovery rate.  Given
a desired false discovery rate, here $10^{-3}$, we can
then identify for which phenomenological parameters
we expect a discovery on average, shown here in grey.}
\label{fig:bayes_disc_sens}
\end{figure}

By sampling from various conditional priors we can 
also quantify how the false and true discovery rates,
or even the distribution of $P$ itself, varies with 
respect to various parameters.  This allows us to 
visualize the sensitivity of the experiment similar 
to Figure \ref{fig:freq_disc_sens}, only for arbitrarily 
complicated models (Figure \ref{fig:bayes_disc_sens}).

\subsection{Predictive Scores}

Lastly we can select a model and claim discovery or no
discovery by comparing the \emph{predictive} performance
of the possible hypotheses.  Here we use our inferences 
to construct a predictive distribution for new data and 
then select the model whose predictive distribution is 
closest to the true data generating process.

Predictive distributions arise naturally in many forms
of inference.  For example, the model configuration 
identified by a frequentist point estimator defines the 
predictive distribution 
\begin{equation*}
\pi_{P} = \pi( y \mid \hat{\theta}(\tilde{y})).
\end{equation*}
Bayesian inference immediately yields two predictive
distributions: the \emph{prior predictive distribution},
\begin{equation*}
\pi_{P} = \pi(y) = \int_{\Theta} \mathrm{d} \theta \,
\pi(y \mid \theta) \, \pi(\theta),
\end{equation*}
and the \emph{posterior predictive distribuiton},
\begin{equation*}
\pi_{P} = \pi(y \mid \tilde{y}) = 
\int_{\Theta} \mathrm{d} \theta \,
\pi(y \mid \theta) \, \pi(\theta \mid \tilde{y}).
\end{equation*}

Regardless of how a predictive distribution is constructed,
it's similarity to the true data generating process, 
$\pi^{*}$, is defined by the Kullback-Leibler divergence,
\begin{align*}
\mathrm{KL} ( \pi_{P} \mid\mid \pi^{*} )
&=
\int_{Y} \mathrm{d} y \, \pi^{*} (y) 
\log \frac{ \pi^{*} (y) }{ \pi_{P} (y) }
\\
&=
\int_{Y} \mathrm{d} y \, \pi^{*} (y) \log \pi^{*} (y)
- \int_{Y} \mathrm{d} y \, \pi^{*} (y) \log \pi_{P} (y).
\end{align*}
Because the first term is the same for all models, the 
relative predictive performance between models is 
quantified by the \emph{predictive score},
\begin{equation*}
S = - \int_{Y} \mathrm{d} y \, \pi^{*} (y) \log \pi_{P} (y).
\end{equation*}

This expectation with respect to the true data generating
process cannot be calculated without already knowing the 
true data generating process, but predictive scores can be 
approximated using observations which, by construction, 
are drawn from that distribution.  Different approximation 
methods combined with various predictive distributions yield 
a host of predictive model comparison techniques, ranging
from cross validation to the Akaike Information Criterion,
to the Bayesian Information Criterion, to Bayesian cross 
validation, the Widely Applicable Information Criterion,
and the Deviance Information Criterion \citep{Betancourt:2015b}.

Error in these approximations, however, can be quite
large and difficult to quantify in practice, leading to
poorly calibrated selection between the absence model 
and the presence model.  Consequently it is critically 
important to estimate their expected false discovery
rate and true discovery rate using the assumed model.
In the frequentist settings these rates can be quantified 
as the minimal performance across the model configurations 
in the absence and presence models, where as in the 
Bayesian setting these rates can be quantified by their
expected performance over the Bayesian joint distribution.

\section{Applications to Limit Setting}

Limit setting is a complement to claiming discovery 
when the experiment is not expected to be sufficiently 
sensitive to the relevant phenomenon.  Instead of claiming
a discovery we consider how strongly we can constrain the 
magnitude of that phenomenon and calibrate the corresponding 
constraint with respect to the absence model.  Because
limit setting is derived from standard inference methods, 
its implementation is significantly more straightforward 
than the implementation of discovery claims.

\subsection{Frequentist Limit Setting with Anchored Confidence Intervals}

In the frequentist setting we can constrain the 
magnitude of a phenomenon by constructing confidence 
intervals than span from a vanishing phenomenon to some
upper limit.  More formally, if the magnitude of the 
phenomenon is positive, so that the absence model is 
defined by $\vartheta = 0$ and the absence model is 
defined by $\vartheta > 0$, then we construct an
\emph{anchored confidence interval} of the form 
$\big[0, \hat{\vartheta}(y) \big)$ that has a 
given coverage, $\alpha$, with respect to the full 
model.  

Given an observation, $\tilde{y}$, we then claim that 
$\vartheta < \hat{\vartheta}(\tilde{y})$ with confidence 
$\alpha$.  The sensitivity of this claim is defined 
with respect to the possible distribution of 
$\hat{\vartheta}(y)$ with respect to the data generating 
process in the absence model. For one and two-dimensional 
absence models the sensitivity can be visualized using 
the same techniques in Section \ref{point_null_complex_alt}.

\subsection{Bayesian Limit Setting with Posterior Quantiles}

The frequentist approach to limit setting has an
immediate Bayesian analogue where we use posterior
quantiles to bound the magnitude of the phenomenon.

For a given \emph{credibility}, $\alpha$, we define
the upper limit, $\vartheta_{\alpha}$ as
\begin{equation*}
\vartheta_{\alpha} (\tilde{y} )
= \big\{ \vartheta \mid 
\int_{0}^{\vartheta} \mathrm{d} \vartheta \,
\pi(\vartheta \mid \tilde{y} ) = 1 - \alpha \big\}.
\end{equation*}
By defining the limit in terms of the marginal
posterior for the phenomenological parameters,
$\vartheta$, we automatically incorporate the
uncertainty in any nuisance parameters into the
bound.

The corresponding sensitivity follows by considering
the distribution $\vartheta_{\alpha}$ with respect
to the Bayesian joint distribution for the absence model.
\begin{enumerate}
\item We first sample a true model configuration from
the absence model by sampling the nuisance parameters
from the conditional prior distribution,
\begin{align*}
\tilde{\sigma} &\sim \pi(\sigma \mid \vartheta = 0)
\\
\tilde{\theta} &= \left(0, \tilde{\sigma} \right).
\end{align*}
\item Next we sample an observation from that model 
configuration,
\begin{equation*}
\tilde{y} \sim \pi(y \mid \tilde{\theta}),
\end{equation*}
\item Finally we compute the inferred upper bound,
\begin{equation*}
\vartheta_{\alpha} (\tilde{y} )
= \big\{ \vartheta \mid 
\int_{0}^{\vartheta} \mathrm{d} \vartheta \,
\pi(\vartheta \mid \tilde{y} ) = 1 - \alpha \big\}.
\end{equation*}
\end{enumerate}

\section{Conclusions and Future Directions}

Both the frequentist and Bayesian perspectives admit 
procedures for analyzing sensitivities and calibrating 
decision making processes.  Implementing these calibrations
in practice, however, is far from trivial.

Frequentist calibration requires bounding the expectation
of a given loss function over all of the data generating 
processes in a given model, or partitions thereof.  
The derivation of analytic bounds from assumptions 
about the structure of the model configuration space and 
the loss function, especially those derived from
asymptotic analyses, is greatly facilitated with the 
presumption of simple model configuration spaces. Numerical 
methods for computing the bounds are also aided by simple 
models.  The probabilistic computations required of Bayesian 
calibration are often more straightforward to approximate but 
sufficiently complex models will eventually frustrate even 
the most advanced Bayesian computational methods.  In practice 
these computational challenges result in a dangerous tension 
between models that are simple enough to admit accurate 
calibrations and models that are complex enough for their
resulting calibrations to be relevant to the experiment 
being analyzed.

The continued improvement in computational resources and
algorithms has gradually reduced, and promises to continue
to reduce, this tension.  Monte Carlo and Markov chain
Monte Carlo methods, for example, have revolutionized
our ability to compute expected losses and Bayesian posterior 
expectations over high-dimensional measurement and model 
configuration spaces.  Unfortunately the applicability of 
this method critically depends on the desired calibration.  
In particular, the square root convergence of Monte Carlo 
estimators is often too slow to ensure accurate calibration 
of rare observations.  This frustrates the calculation, for 
example, of the $\mathcal{O}(1 - 10^{-7})$ significance 
thresholds presumed in contemporary particle physics.  
Computational limitations restrain not only the complexity 
of our models but also the complexity of the loss function
we consider.

Statistics is a constant battle between computational
feasibility and compatibility with analysis goals.  
Ultimately it is up to the practitioner to exploit their 
domain expertise to identify compromises that facilitate 
high-performance decision making.

Finally there is the issue of experimental design
where we tune the design of an experiment to achieve
a given performance.  As difficult as it is to compute
this performance, ``inverting'' the calibration to
identify the optimal experiment is even harder.
For complex models that don't admit analytic results,
contemporary best practice often reduces to exploring 
the experiment design space heuristically, guided by
computed calibrations and domain expertise.  

An interesting future direction is the use of automatic 
differentiation methods to automatically estimate not 
only the expected losses but also their gradients with 
respect to the experimental design.  Although an imposing
implementation challenge, these gradients have the
potential to drastically improve the exploration and
optimization of experimental designs.

\section{Acknowledgements}

I thank Lindley Winslow, Charles Margossian, Joe Formaggio,
and Dan Simpsons for helpful comments and discussions.

\bibliography{experimental_sensitivity}
\bibliographystyle{imsart-nameyear}

\end{document}